%% file: bare_conf_NDSS2025.tex
\documentclass[conference]{IEEEtran}
\ifCLASSINFOpdf
\else
\fi
\usepackage{multirow}
\usepackage{multicol}
\usepackage{booktabs}
\usepackage{adjustbox,mdframed}
\usepackage{balance}
\usepackage{amsthm}
\usepackage{amsmath}
\usepackage{pifont}
\usepackage{amsmath}
\usepackage{fontawesome}
\usepackage{makecell}
\usepackage{url}
\usepackage{xcolor,colortbl}

\usepackage{caption}
\theoremstyle{definition}

\usepackage{multirow}
\usepackage{multicol}
\usepackage{booktabs}
\usepackage{adjustbox}
\usepackage{amsthm}
\usepackage{amsmath}
\usepackage{pifont}
\usepackage{amsmath}
\usepackage{fontawesome}
\usepackage{makecell}
\usepackage{marginnote}

\usepackage{url}

\usepackage{breakurl}
\usepackage[breaklinks]{hyperref}

\usepackage{xcolor,colortbl}
\usepackage{graphicx}
\usepackage[normalem]{ulem}

\usepackage{caption}
\theoremstyle{definition}

\newcommand{\model}[0] {\textit{Tweezers}}
\newcommand{\ra}[1]{\renewcommand{\arraystretch}{#1}}

\newif\ifdiff

\difftrue

\begin{document}
\title{\model{}: A Framework for Security Event Detection via Event Attribution-centric Tweet Embedding}

\author{
    \IEEEauthorblockN{
        Jian Cui\IEEEauthorrefmark{1}\textsuperscript{\textsection},
        Hanna Kim\IEEEauthorrefmark{2},
        Eugene Jang\IEEEauthorrefmark{3},
        Dayeon Yim\IEEEauthorrefmark{3},
        Kicheol Kim\IEEEauthorrefmark{3},
        Yongjae Lee\IEEEauthorrefmark{3},
        Jin-Woo Chung\IEEEauthorrefmark{3}, \\
        Seungwon Shin\IEEEauthorrefmark{2},
        Xiaojing Liao\IEEEauthorrefmark{1}
    }
    \IEEEauthorblockA{
    \IEEEauthorrefmark{1}Indiana University Bloomington, \IEEEauthorrefmark{2}KAIST, \IEEEauthorrefmark{3}S2W Inc.}
    
    \IEEEauthorblockA{\IEEEauthorrefmark{1}\{cuijian, xliao\}@iu.edu, \IEEEauthorrefmark{2}\{gkssk3654, claude\}@kaist.ac.kr, \IEEEauthorrefmark{3}\{genesith, dayeon, kkim, lee, jwchung\}@s2w.inc}
}


\IEEEoverridecommandlockouts
\makeatletter\def\@IEEEpubidpullup{6.5\baselineskip}\makeatother
\IEEEpubid{\parbox{\columnwidth}{
		Network and Distributed System Security (NDSS) Symposium 2025\\
		23–28 February 2025, San Diego, CA, USA\\
		ISBN 979-8-9894372-8-3\\
		https://dx.doi.org/10.14722/ndss.2025.23139\\
		www.ndss-symposium.org
}
\hspace{\columnsep}\makebox[\columnwidth]{}}

\maketitle

\begingroup\renewcommand\thefootnote{\textsection}
\footnotetext{Work performed while at S2W Inc.}
\endgroup

\input{sections/0_abstract}


% \fi
\IEEEpeerreviewmaketitle

\input{sections/1_introduction}

\input{sections/2_background}
\input{sections/3_method}

\input{sections/4_framework}
\input{sections/4_evaluation}
\input{sections/5_usecases}

\input{sections/6_discussion}

\input{sections/7_relatedwork}
\input{sections/8_conclusion}

\section*{Acknowledgment}
We appreciate the reviewers’ valuable and constructive feedback. 
This work was supported by Institute of Information \& communications Technology Planning \& Evaluation (IITP) grant funded by the Korea government (MSIT) [RS- 2023-00215700, Trustworthy Metaverse: blockchain-enabled convergence research].
Xiaojing Liao is supported in part by the National Science Foundation (CNS-1850725, 2343618) and the Luddy Faculty Fellowship.

\bibliographystyle{plain}
\bibliography{references}
\input{sections/appendix}
\end{document}

%% file: sections/0_abstract.tex
\begin{abstract}
Twitter is recognized as a crucial platform for the dissemination and gathering of Cyber Threat Intelligence (CTI). Its capability to provide real-time, actionable intelligence makes it an indispensable tool for detecting security events, helping security professionals cope with ever-growing threats. However, the large volume of tweets and inherent noises of human-crafted tweets pose significant challenges in accurately identifying security events. While many studies tried to filter out event-related tweets based on keywords, they are not effective due to their limitation in understanding the semantics of tweets. Another challenge in security event detection from Twitter is the comprehensive coverage of security events. Previous studies emphasized the importance of early detection of security events, but they overlooked the importance of event coverage. To cope with these challenges, in our study, we introduce a novel event attribution-centric tweet embedding method to enable the high precision and coverage of events. Our experiment result shows that the proposed method outperforms existing text and graph-based tweet embedding methods in identifying security events. Leveraging this novel embedding approach, we have developed and implemented a framework, \textit{Tweezers}, that is applicable to security event detection from Twitter for CTI gathering. This framework has demonstrated its effectiveness, detecting twice as many events compared to established baselines. Additionally, we have showcased two applications, built on \textit{Tweezers} for the integration and inspection of security events, i.e., security event trend analysis and informative security user identification. 
\end{abstract}

%% file: sections/1_introduction.tex
\section{Introduction}
Social networks, especially Twitter (now known as X), have become invaluable resources for security practitioners by enabling the sharing of up-to-date security information and the collection of actionable cyber threat intelligence (CTI) information. 
Twitter's large user base and real-time communication capabilities facilitate its use in vulnerability monitoring and threat intelligence gathering, as evidenced in a report by Trend Micro~\cite{trendmicro_twitter}.
Furthermore, Twitter has been integrated into the OSINT (Open Source Threat Intelligence) tools developed by Recorded Future~\cite{recorded_osint}.
Prior research has also demonstrated the Twitter's utility in security applications, including vulnerability disclosure~\cite{sabottke2015vulnerability}, IOC gathering~\cite{twiti, icominer}, DDoS attack forecasting~\cite{wang2017ddos}, and security event detection~\cite{shin2020cybersecurity, le2017sonar, cydec}.

Security event detection on platforms like Twitter is essential for the proactive management of cyber risks.
A \emph{security event} is a specific occurrence or incident that may affect the confidentiality, integrity, or availability of an organization’s information technology (IT) systems, data, or overall cybersecurity posture.
Notable instances of such events reported on Twitter in November 2022 include the Medibank data leak~\cite{medibank_leak}, OpenSSL vulnerability disclosure~\cite{openssl}, phishing campaign in TikTok~\cite{tiktok_invisible}, etc.
The continual evolution and complexity of threats make it imperative for security professionals to stay informed about such events. 
By leveraging Twitter's real-time data, cybersecurity professionals can access immediate and actionable intelligence, which is critical for timely responses to threats such as phishing campaigns, vulnerability exploits, etc. 
Although security events are also available in traditional news outlets, they fail to provide comprehensive actionable threat-intelligent information.
Additionally, studies~\cite{cydec, shin2020cybersecurity} have shown that Twitter's timely reporting of security events frequently precedes traditional media, thereby providing a more immediate context for assessing and responding to cyber threats (details in Section~\ref{sec:background}).

While the importance of extracting security events from Twitter is widely acknowledged, the sheer volume and inherent noise in tweets pose significant challenges to security event detection. 
To filter out tweets about security events, existing work employs different techniques to identify the security event-related keywords (e.g., malware name, vulnerability ID, hacking group) in tweets~\cite{shin2020cybersecurity, le2017sonar, khandpur2017crowdsourcing}.
For instance, W2E~\cite{shin2020cybersecurity} tries to find security event-related keywords by monitoring new and re-emerging security keywords from tweets.
Similarly, SONAR~\cite{le2017sonar} develops a GloVe-based keyword-finding module to identify event-related keywords.
However, these keyword-based methods often lead to irrelevant tweets due to the ambiguity of keywords.
For example, in our analysis, W2E identified ``Tropic'' as a keyword, which is linked to an event where the ``Tropic Trooper'' hacking group distributed the SMS Bomber Tool.
However, utilizing this keyword to filter for security-related tweets also brought up non-security event-related tweets discussing ``Tropic Thunder'' or ``Tropic Rush.'' 
It indicates the fundamental issue with these keyword-based approaches, as they cannot consider tweets' semantics. 
Although there are many semantic-aware text embedding methods, such as BERT~\cite{bert}, RoBERTa~\cite{roberta}, etc, have been proposed recently, they are not effective in distinguishing event-related tweets. 
Specifically, in the security domain, different events often encompass similar topics, leading these embedding methods to erroneously suggest similarities between tweets from distinct events (See Section~\ref{sec:method_overview} for detailed explanation).

Another challenge in security event detection is the timely and comprehensive coverage of events, as it helps security practitioners gain insights into different types of threats, tactics, and vulnerabilities that may be relevant to their organization.
While previous research~\cite{shin2020cybersecurity, sabottke2015vulnerability} has primarily emphasized the importance of early detection of security events, it has often overlooked the importance of event coverage.
Given that limited coverage can result in missing important events, comprehensive coverage of security events is as important as early detection of events.
According to our experiment in Section~\ref{sec:exp_wild}, prior work only achieves 2.7\% and 29.8\% of event detection coverage. 
These approaches typically assume that tweets related to various events exhibit distinct lexical and syntactic patterns. 
However, in the context of security event detection, it is common for different events, such as various vulnerability exploitations or the emergence of new malware strains, to exhibit similar terminologies and syntactic structures in tweets. 
Therefore, to effectively distinguish these events and thus improve the coverage of the detection, there is a need for specialized approaches tailored for detecting security events on platforms like Twitter.

In this study, we introduce security event attribution-centric embedding, a novel method for generating tweet embeddings for security event detection.
Utilizing clustering algorithms on these embeddings enables us to detect security events with high precision and comprehensive coverage.
Our tweet embedding approach begins by extracting security event-related entities from each tweet, leveraging semantics through advanced Named Entity Recognition (NER) methods based on Large Language Models (LLMs).
The underlying assumption is that tweets concerning the same event will contain the same entities. 
To effectively represent tweets with shared entities, we construct a \textit{tweet relation graph} by connecting tweets with the same entities.
This graph forms the basis of a Graph Attention Network (GAT)~\cite{gatv2}, providing insights into tweets that have common security-related entities.
Moreover, we enhance the embedding quality by integrating additional event-related information, including tweet content, security categories, and temporal data. 
This data is vectorized and constitutes the initial feature set for each node within the tweet relation graph.
With the tweet relation graph and event-related information, the GAT is optimized through a specialized function designed for event detection tasks.
Based on this method, we have also developed a framework, \model{}, which is specifically designed for detecting security events from a stream of tweets.

Moreover, \model{} extends its capabilities to analyze security event trends and identify informative security users on Twitter. 
Through analysis, the framework not only highlights the prominent events, such as significant vulnerabilities and data breaches but also provides insights into the evolving nature of threats,
Additionally, \model{} helps pinpoint Twitter users who offer in-depth security insights and possess extensive influence, thereby improving early threat detection capabilities. 
Compared to traditional methods that only quantify security-related posts, our method can find Twitter users capable of offering detailed event analyses. 

To summarize, the major contributions of our work are:
\begin{itemize}
    \item We propose a novel event attribution-centric tweet embedding method, which is specifically tailored for security event detection. Our evaluation shows that our embedding surpasses existing text and graph-based tweet embedding methods in security event identification.
    \item With our event attribution-centric tweet embedding method, we design an event detection framework, \model{}, that can be used in real-world scenarios. \model{} is tested with real-world data and performs significantly better than existing baselines regarding detection precision and event coverage. 
    \item We provide two practical use cases to showcase the practical applications of our framework. These use cases involve analyzing security event trends and identifying informative security users.
    \item We release the code and dataset for training and evaluating our event attribution-centric embedding\footnote{https://github.com/jiancui-research/tweezers}. Additionally, the trained multi-label tweet categorizer employed in our framework will also be made available for future research.
\end{itemize}

%% file: sections/2_background.tex
\section{Background \& Problem Statement}
\label{sec:background}
\subsection{Security Event Detection}

Based on the existing literature~\cite{le2017sonar, shin2020cybersecurity} and the NIST glossary~\cite{security_incident_nist}, our study defines a security event as \emph{a specific occurrence or incident that may affect the confidentiality, integrity, or availability of an organization’s information technology (IT) systems, data, or overall cybersecurity posture.}
In other words, security events in our study encompass a range of occurrences that have implications for cybersecurity.

This study primarily focuses on five security categories: Vulnerability, Ransomware/Malware, Data Privacy, Fraud/Phishing, and DoS/DDoS\footnote{Detailed definitions of each category can be found in our released artifact}.
Examples of security events include the disclosure of an OpenSSL vulnerability~\cite{openssl}, Proxy Bots involvement in DDoS attacks~\cite{proxybot_ddos_event}, and phishing campaigns via Google Play~\cite{phishing_googleplay_event}. 

Detecting such events is crucial as it provides actionable intelligence to security professionals, enabling efficient threat identification and mitigation. 
For instance, identifying a new OpenSSL vulnerability allows experts to apply necessary patches by referencing associated CVEs. 
Similarly, knowledge of Proxy Bot-related events can facilitate the identification of malicious IP addresses and code hashes crucial for defensive operations, and awareness of phishing events can help prevent individuals from falling victim to such schemes with Phishing URLs associated.

\subsection{Twitter as a CTI source}

\noindent \textbf{Advantages of Twitter.}
Social media has been recognized as an important source of threat intelligence (TI) by both the TI industry~\cite{recorded_TI_feed, recorded_osint, trendmicro_twitter} and academia~\cite{shin2020cybersecurity, sabottke2015vulnerability}. 
Although other sources such as blogs, news media, and dark/deep web forums are available, extracting TI from unstructured social media feeds, especially Twitter, presents unique challenges, including large data volumes and data sparsity~\cite{sun2023cyber, ritter2015weakly}. 
Despite these challenges, Twitter offers significant advantages due to its comprehensive and real-time coverage of threat intelligence, making it a valuable platform for security practitioners.

One of the key reasons Twitter stands out is its role as a dynamic hub where diverse security practitioners share information, providing a broader spectrum of threat intelligence. 
For instance, different tweets about the same WordPress vulnerability might provide various details—one might list IP addresses involved in the attack~\cite{wordpress_tweet_1}, offering specific data points for analysis, while another might discuss associated CVEs and attack methods~\cite{wordpress_tweet_2}.

Moreover, Twitter often reports on security events faster than traditional media, as demonstrated by previous studies~\cite{shin2020cybersecurity, cydec} and practical observations.
For instance, the early reporting of the MacOS RustDoor backdoor on Twitter~\cite{macos_backdoor_twitter} occurred two days before its mention on TheHackerNews~\cite{macos_backdoor_news}.

In summary, Twitter is a vital part of modern threat intelligence frameworks, offering timely and actionable intelligence for effective cybersecurity. 
Monitoring Twitter allows security professionals to detect and respond to threats in real time, significantly enhancing the capability to manage cybersecurity risks.
Thus, the goal of this paper is to address the challenges of data volume and sparsity issues on Twitter, aiding the efficient use of Twitter data in TI by effectively detecting security events.

\noindent \textbf{Twitter vs. other sources}
In addition to social media, multiple sources can be used for threat intelligence, such as blogs and dark/deep web forums. 
However, prior studies have shown that social media, especially Twitter, is chosen as the better source for open-source intelligence by security practitioners due to its widespread use and the valuable CTI information it provides for further analysis~\cite{tundis2020automated}.

Although third-party websites such as blogs sometimes provide more detailed analysis, Twitter is considered as a much more timely source; also, these sources are often shared on Twitter with external links~\cite{turkish_espionage, no_justice_wiper}, making it easier to check associations with related posts for additional details. 
Moreover, other sources may not cover as comprehensive topics as Twitter. 
For instance, the dark web is rich in information related to cybercrime, such as drug and sex trafficking, but lacks extensive details on topics like DDoS attacks and phishing~\cite{nazah2020evolution}. 
Similarly, hacking forums are more concentrated on data breaches and vulnerability exploits~\cite{top_deep_forums}.

On the other hand, social media, especially Twitter, has been an indispensable platform for understanding and analyzing a broad range of security topics such as malware discovery~\cite{malware_twitter}, Indicators of Compromise (IOC) extraction~\cite{twiti}, vulnerability analysis~\cite{sabottke2015vulnerability}, phishing detection~\cite{kim2023drainclog}, and more.

\subsection{Problem Statement}
Given a stream of security-related tweets, denoted as $\mathcal{T} = \{t_1, t_2, \ldots\}$, our primary objective is the identification of security event sets $\mathcal{E} = \{e_1, e_2, \ldots\}$, where each event $e_i$ is a subset of the tweet collection $\mathcal{T}$, comprising tweets that pertain to a specific event.
In other words, the task of event detection can be expressed as the determination of highly related tweet sets $\mathcal{E}$ within the given set of tweets $\mathcal{T}$, mathematically represented as:
$$\mathcal{E} = f (\mathcal{T})$$

%% file: sections/3_method.tex
\begin{figure}[t!]
    \centering
    \includegraphics[width=\columnwidth]{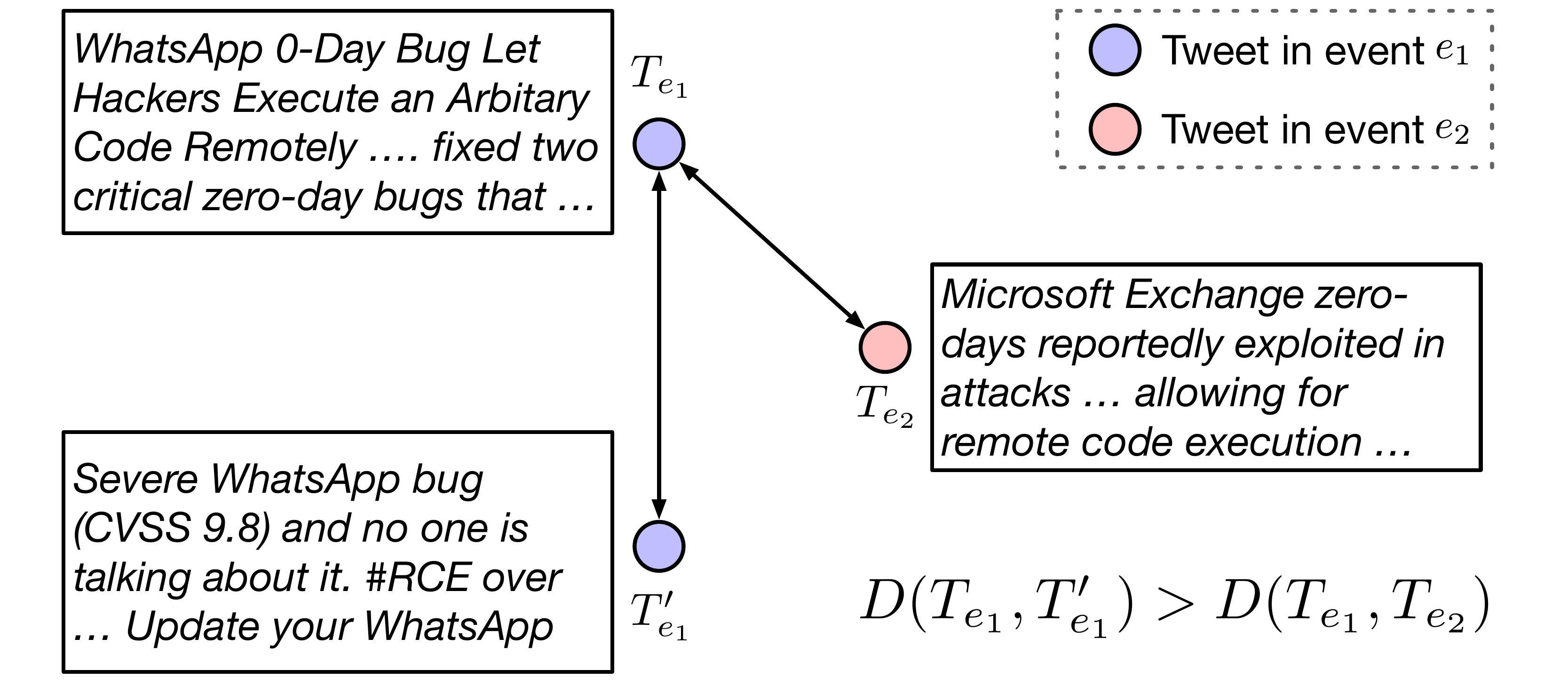}
    \caption{Tweets embedded with Word2Vec. The distance between embeddings of tweets belonging to the same event is larger than those belonging to different events.}
    \label{fig:distance_embedding}
    \vspace{-5pt}
\end{figure}

\section{Event Attribution-centric Tweet Embedding}
\label{sec:emb_method}
This section provides an overview of our tweet embedding generation method, security event attribution-centric tweet embedding, followed by detailed descriptions of each step involved.

\subsection{Method Overview}
\label{sec:method_overview}

While previous studies~\cite{bert_event_detection, word2vec_event_detection} have leveraged text embedding methods, such as Word2Vec~\cite{word2vec} or GloVe~\cite{glove}, for event detection, they operate under the assumption that distinct events involve different lexical or syntactic features. 
However, in the security domain, it is notable that different events can share similar topic terms and syntactic structures.
For instance, different incidents involving the exploitation of a zero-day vulnerability may exhibit overlapping topic terms (e.g., zero-day, exploitation) and syntactic similarities, as illustrated by the sample tweets shown in Figure~\ref{fig:distance_embedding}. 
Tweet $T_{e_1}$ and $T_{e_2}$, although related to different events, $e_1$ and $e_2$, show similar syntactic structures and topic terms. 
As a result, text embedding methods like the Word2Vec, can falsely suggest the similarities between tweets from distinct events. 
As shown in Figure~\ref{fig:distance_embedding}, the proximity between tweets, $T_{e_1}$ and $T_{e_2}$, which are from different events, is closer than that of tweets from the same events, $T_{e_1}$ and $T'_{e_1}$.   
This means that relying solely on current text embedding techniques is insufficient for effective security event detection (see Section~\ref{sec:exp_emebdding}).

\begin{figure}[t!]
    \centering
    \includegraphics[width=.8\columnwidth]{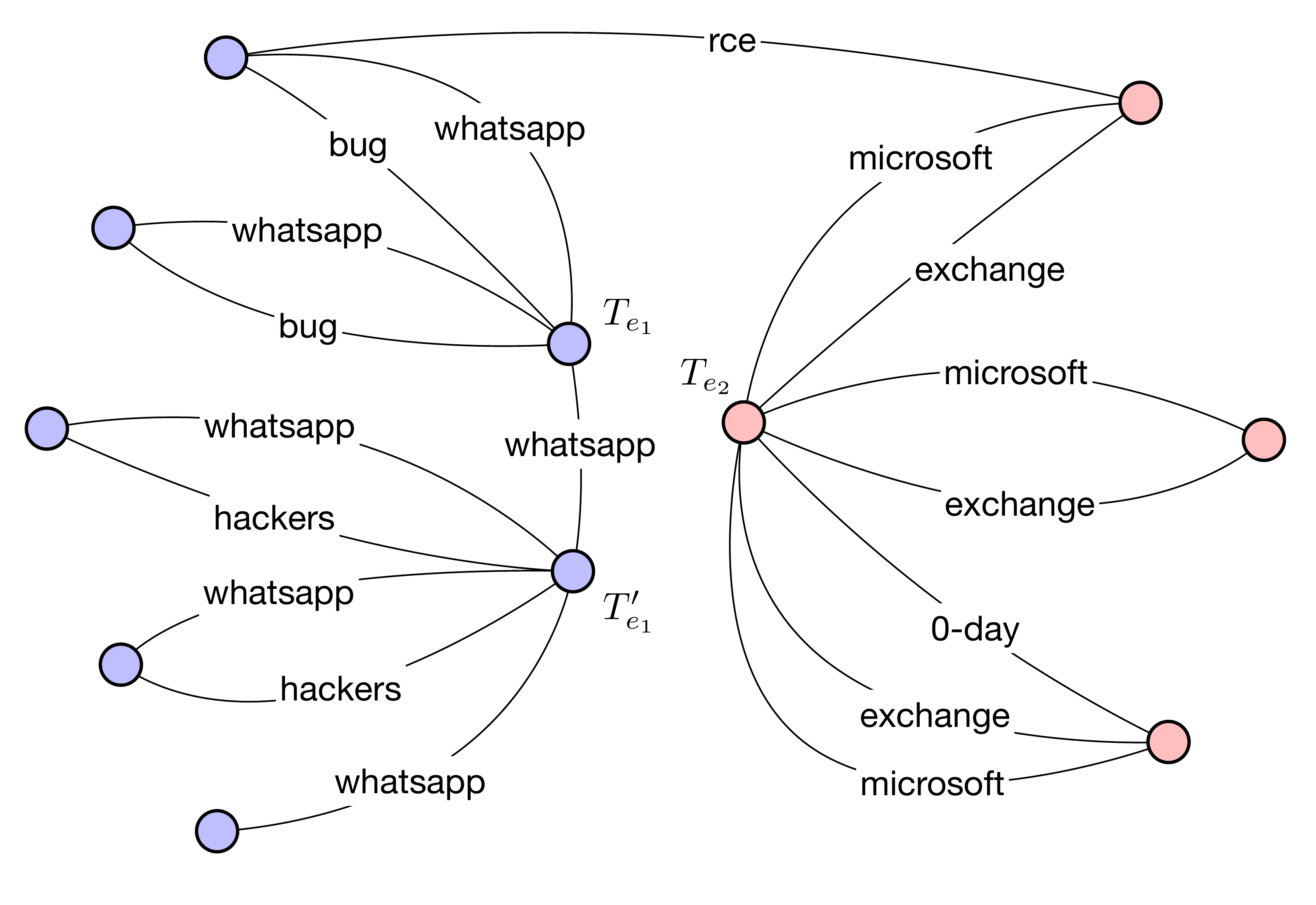}
    \caption{In the tweet relation graph, the one-hop neighbors of tweets $T_{e_1}$, $T'_{e_1}$, and $T_{e_2}$—referenced in Figure~\ref{fig:distance_embedding}—are illustrated. Purple and pink dots represent tweets associated with the corresponding events.}
    \label{fig:graph_viz}
    \vspace{-5pt}
\end{figure}

To effectively differentiate between different security events, it is crucial to identify critical attributions for security events, such as threat actors, victim organizations/individuals, attack patterns, activities, methods, etc.
These details provide a precise description of specific aspects of a security event, so identifying tweets that share such information is crucial to distinguishing different security events.
To this end, we explore the event attribution related to cybersecurity. Specifically, we extend the utilization of security event entities within the Structured Threat Information Expression (STIX) standard~\cite{stix}, a well-established protocol in the realm of Cyber threat intelligence. 
In our study, we selected 13 entities relevant to security event attribution.

\begin{figure}
    \centering
    \includegraphics[width=\columnwidth]{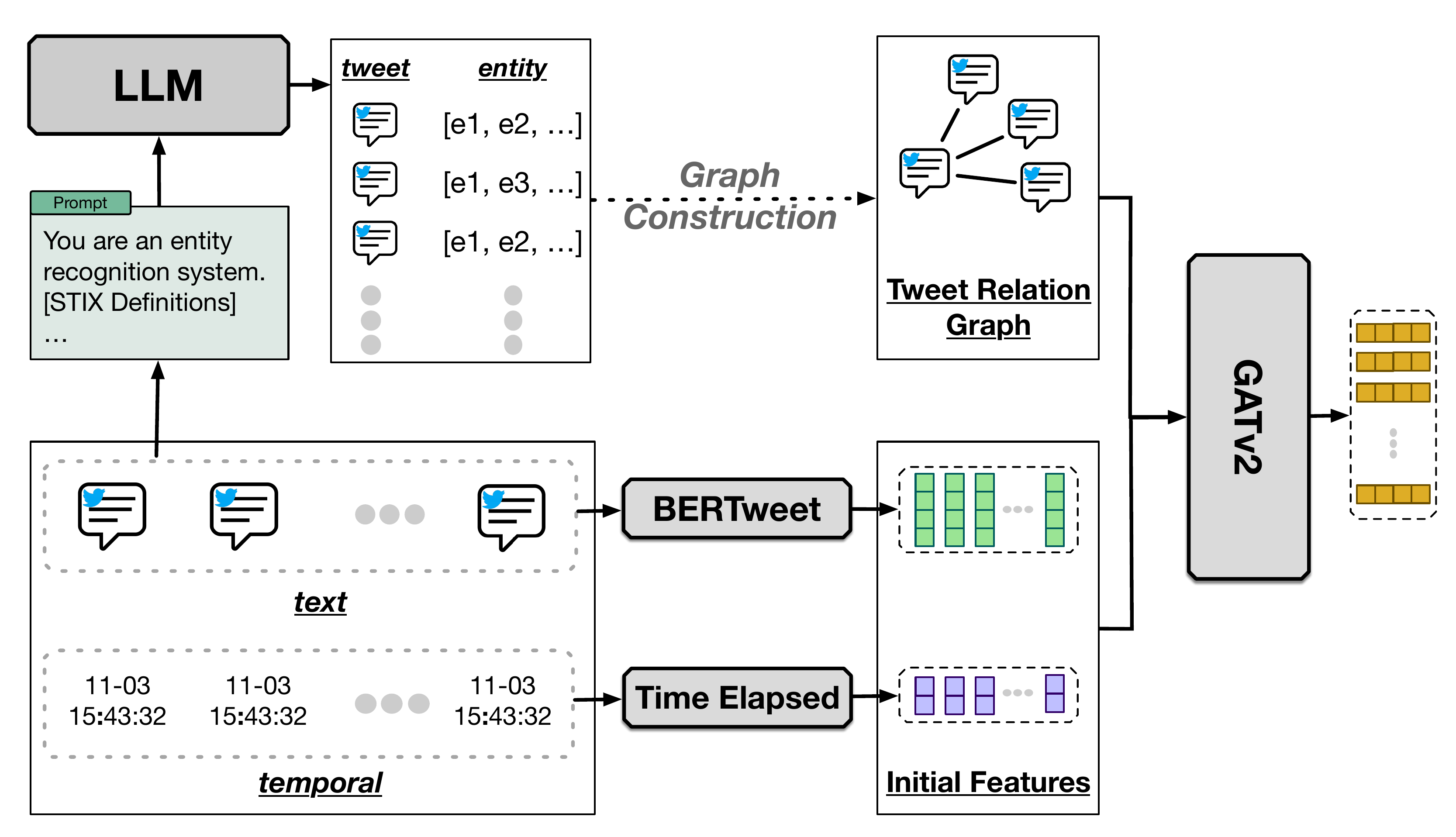}
    \caption{Overview of security event attribution-centric tweet embedding method.}
    \label{fig:emb_generator}
    \vspace{-5pt}
\end{figure}

After that, we incorporate these event attributions into tweet embeddings for effective security event detection.
We argue that graph-based tweet embedding preserves rich relationships and event attributions among tweets while robustly handling sparse security information, serving as valuable evidence to distinguish between tweets related to different events.
In our study, we construct a tweet relation graph to generate tweet embedding, where each node represents an individual tweet and connections between nodes are established based on the shared security entities.
Particularly, we employ a Graph Attention Network (GAT), a deep learning architecture that processes graph-structure information.
The GAT takes input from the tweet relation graph along with a series of node features, such as tweet content and temporal information. 
GAT is trained with dedicated objective functions, ensuring the effective integration of all relevant information into the final output tweet embeddings.
After training, the GAT can integrate entity-sharing security event attributions presented in the tweet relation graph to generate effective embeddings for event detection. 
The overview of this approach is shown in Figure~\ref{fig:emb_generator}.

Figure~\ref{fig:graph_viz} shows the one-hop neighbors of the tweets mentioned in Figure~\ref{fig:distance_embedding}.
Compared with Figure~\ref{fig:distance_embedding}, it shows a clearer differentiation between the security events of ``WhatsApp vulnerability exploitation'' and ``Microsoft Exchange vulnerability'' events.
It is worth noting that tweets belonging to event $e_1$ are mainly connected to other tweets within the same event since WhatsApp, the victim organizations are extracted from tweets belong event $e_1$. 
Meanwhile, tweets pertaining to event $e_2$ are connected predominantly to other tweets within event $e_2$, highlighting the distinct nature of the two events.

\subsection{Tweet Embedding}
\label{sec:embedding_generator}

We provide details on the construction of the tweet relation graph, and the derivation of tweet-related features, followed by a detailed explanation of our embedding generation methodology.

\noindent\textbf{Tweet relation graph construction}.
As mentioned earlier, the Tweet Relation Graph (TRG), which represents shared security event attributions among tweets, is constructed to generate tweet embeddings for security event detection. 
The TRG can be described as
$\mathcal{G} = (\mathcal{V}, \mathcal{E})$, where $\mathcal{V}$ represents the nodes (tweets), and $\mathcal{E}$ represents the edges in the graph.
The tweet relation graph is constructed so that edges connect any two tweets that share the same entities related to security event attribution.
From the 18 STIX Domain Objects (SDOs) outlined in STIX 2.1~\cite{stix}, we select 13 entity types relevant for security event attribution (excluding non-essential entity types such as Report, Note, Observed data, etc).

The process of extracting these entities from unstructured text is commonly known as Named Entity Recognition (NER) in the Natural Language Processing (NLP) domain. 
While numerous deep learning models have been proposed and proven effective in NER tasks, they rely on human labor to annotate the predefined entities for training these NER models.
However, the success of generative Large Language Models (LLMs) has simplified the NER task to work in zero-shot settings. 
As indicated by recent studies~\cite{ashok2023promptner}, prompt engineering on generative LLMs shows superior performance in NER tasks compared to traditional NER models.
To assess the effectiveness of prompt-based NER within a security context, we conduct a preliminary experiment with a cybersecurity NER dataset, CyNER~\cite{cyner}.
The results reveal that while prompt-based NER does not outperform all trained NER models, it surpasses the RoBERTa-based model (RoBERTa-base recording an F1 score of 39.7, compared to 50.19 for prompt-based NER), even without any training.
Comprehensive details of the experiment are provided in the Appendix~\ref{sec:ner_eval}.
Aligned with the methodology proposed by previous research~\cite{ashok2023promptner}, we incorporate definitions for 13 security-related entities in the prompt.
Subsequently, the extracted entities are used in the construction of the tweet relation graph.

\noindent\textbf{Tweet-related feature engineering}. 
We identify three critical pieces of information as TRG's node features: tweet content, security category, and temporal information. 
The encoding details for each piece of information are elaborated below. 
The resulting encoded vectors are then concatenated to form the initial features of the corresponding nodes in the tweet relation graph.

\begin{enumerate}
    \item \textbf{Tweet content}: The (text) content of the tweet contains information indicative of the discussed event.
    The tweet content is converted into a 768-dimensional vector by the BERTweet model, which was chosen for its specialization in the Twitter domain.
    As noted previously, embeddings of tweet text alone cannot provide a comprehensive representation of a tweet's information.
    
    \item \textbf{Temporal information}:
    Tweets describing the same event tend to be posted within a similar timeframe.
    The temporal information of tweets is utilized as two-dimensional features (hours and days elapsed since the beginning of 2020).
\end{enumerate}

\noindent\textbf{Tweet embedding generation}.
The TRG and its node features are processed with the enhanced version of GAT, GATv2~\cite{gatv2}.
For a node (tweet) $v$, the output embedding $\mathbf{h}'_v$ is generated through the following equations.  
\begin{equation}
\mathbf{h'}_v = \sum_{v' \in N(v)} \alpha \cdot \mathbf{W} \mathbf{h}_{v'}    
\end{equation}
The attention score $\alpha$ is obtained by: 

\begin{equation}
\begin{split}    
    e_{v} &= \mathbf{a}^{T} \cdot LeakyReLU(\mathbf{W} \cdot [\mathbf{h}_v \mathbin\Vert \mathbf{h}_{v'}]) \\
    \alpha  &= softmax(e_{v}) = \frac{exp(e_{v})}{\sum_{v' \in N_v} exp(e_{v'})} \\
\end{split}
\end{equation}
where $\mathbf{h}_v$ and $\mathbf{h}_v'$ are initial features of node $v$ and its neighbor nodes $v'$, respectively. 
$\mathbf{W}$ and $\mathbf{a}$ are learnable weights of the GATv2 model.
Note that GATv2 utilizes attention to learn which neighboring nodes should be more influential.

\begin{figure}[t]
    \centering
    \includegraphics[width=.8\columnwidth]{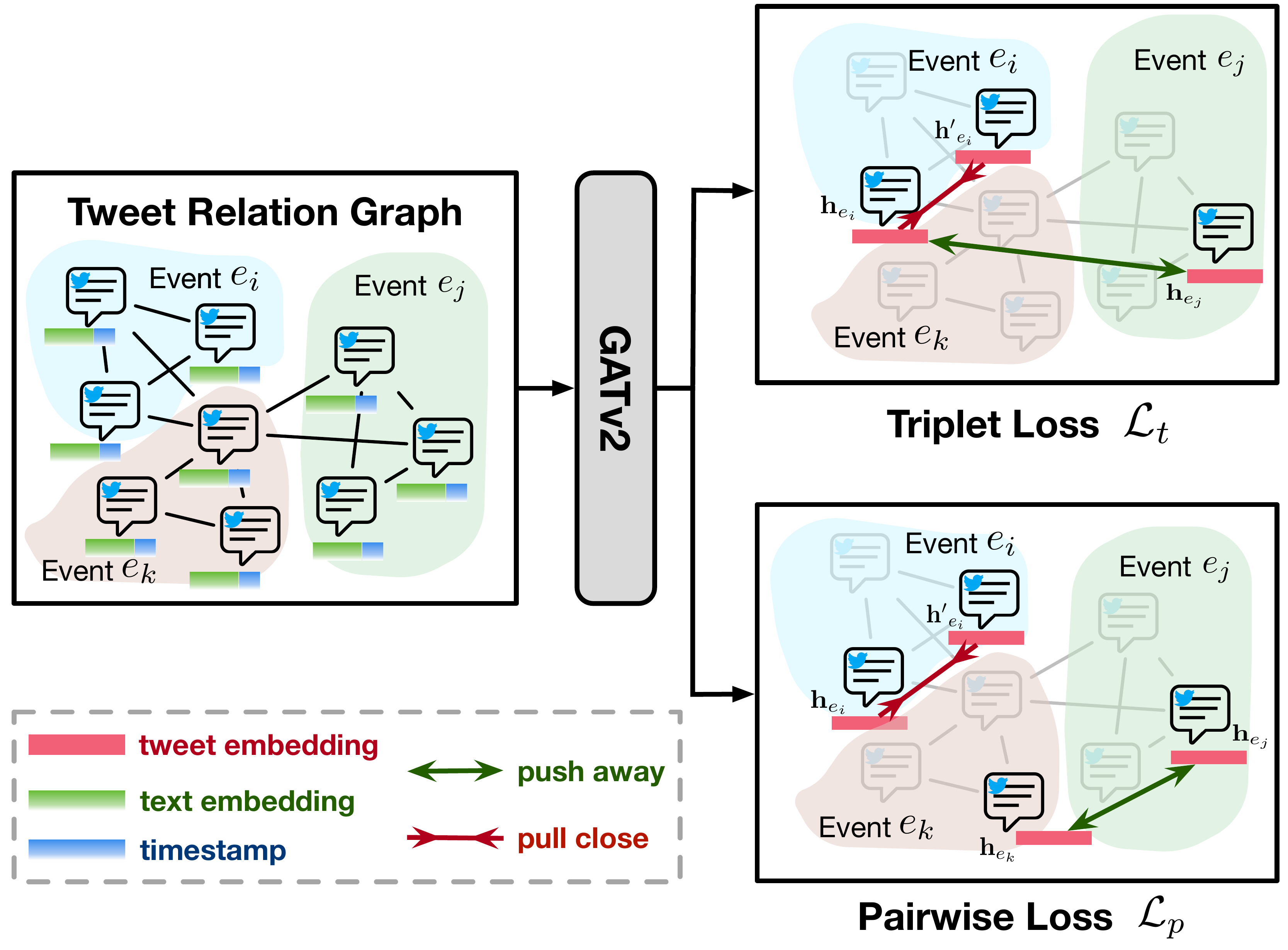}
    \caption{Explanation of our objective function for tweet embedding method. 
    % Triplet loss,  $\mathcal{L}_t$, is to ensure that the distance between tweet embedding $\mathbf{h}_{e_i}$ and tweets belonging to the same event get closer than tweets belonging to different events. Pairwise loss, $\mathcal{L}_p$, is to ensure the tweets belonging to the same event get closer than tweets belonging to different events.
    }
    \label{fig:loss_func}
    \vspace{-5pt}
\end{figure}

To enable security event clustering, embeddings of tweets belonging to the same event should be close to each other, while embeddings of tweets associated with different events should be kept far apart from each other. 
This can be optimized using the contrastive learning technique of triplet loss~\cite{triplet_loss}.
Triplet loss uses an anchor tweet embedding $\mathbf{h}_{e_i}$ and compares its distances to embeddings of a tweet of the same event and embeddings of a tweet of a different event.
The formulation of triple loss in our case is:
\begin{equation}
\mathcal{L}_t = max (\mathbin\Vert \mathbf{h}_{e_i} - \mathbf{h'}_{e_i} \mathbin\Vert - \mathbin\Vert \mathbf{h}_{e_i} - \mathbf{h}_{e_j} \mathbin\Vert + \alpha, 0)
\end{equation}
where $\mathbf{h}_{e_i}$ and $\mathbf{h'}_{e_i}$ refer to embeddings of two tweets belonging to the same event $e_i$ and $\mathbf{h}_{e_j}$ refer to embeddings of a tweet belong to another event, $e_j$.
$\alpha$ represents the margin between positive and negative pairs. 

Another contrastive learning method is to manipulate the distance between tweets on a pairwise basis using the pairwise loss function~\cite{ren2022known}.
The pairwise loss function is: 
\begin{equation}
\mathcal{L}_p = max (\mathbin\Vert \mathbf{h}_{e_i} - \mathbf{h'}_{e_i} \mathbin\Vert - \mathbin\Vert \mathbf{h}_{e_j} - \mathbf{h}_{e_k} \mathbin\Vert + \alpha, 0)    
\end{equation}
where $\mathbf{h}_{e_i}$ and $\mathbf{h'}_{e_i}$ refer to embeddings of two tweets belonging to same event $e_i$, 
while $\mathbf{h}_{e_j}$ and $\mathbf{h}_{e_k}$ refer to embeddings of tweets belonging to separate events $e_j$ and $e_k$, respectively.

Both loss functions are summed up and utilized as the objective function to optimize the learnable parameters of the GATv2 model. 
An intuitive visualization of how these two loss functions operate is shown in Figure~\ref{fig:loss_func}.

\subsection{Effectiveness Analysis}
\label{sec:exp_emebdding}
In this section, we look into our model's capability to capture the semantic similarity of security event tweets and compare it with other tweet embedding techniques, TF.IDF and Word2Vec, employed in existing security event detection frameworks~\cite{shin2020cybersecurity, le2017sonar, kristiansen2020cti}. 
We also compare our embedding method with transformer-based language models, which have been proven to be effective in many NLP tasks and also applied in disaster prediction on Twitter~\cite{bert_event_detection}. 
Moreover, some embedding methods that are proposed for social event detection~\cite{cao2021knowledge, ren2022known} are also compared in this section. 

\noindent \textbf{Dataset}.
In our study, we created a tweet dataset related to 182 cybersecurity events.
We identified noteworthy security events by monitoring three different sources: The Hacker News\footnote{https://www.thehackernews.com}, BleepingComputer\footnote{https://www.bleepingcomputer.com}, and Hackmageddon\footnote{https://www.hackmageddon.com}.
For each event, we collect tweets discussing the event through a manual process of inspecting tweets from the period when the event was reported.

Between June 1, 2022, and October 21, 2022, we identified 138 events. 
To ensure our model’s robustness to changes over time, we also included dataset with the latest tweets from January and February 2024. 
This additional dataset comprises 204 tweets from 21 events in January and 234 tweets from 23 events in February.
In total, we found 2054 relevant tweets from 1,119 different Twitter accounts, averaging 11 tweets per event.
%
% Aligned with Twitter API's policy~\cite{twitter_privacy_policy}, we have released this dataset\footnote{https://github.com/jiancui-research/tweezers.git}, which consists of an event list and corresponding tweet IDs.

\input{tables/clustering_dataset_exp}

\noindent \textbf{Implementation and experimental setup.}
The dataset is partitioned into training, validation, and test sets based on distinct time periods and security events, as detailed in Table~\ref{tab:clustering_dataset_exp}. 
Note that Test datasets do not overlap with the training and validation set, and Test-2 and Test-3 have significant temporal gaps compared to Test-1.
Regarding the implementation of the embedding method, we employ PyTorch and the Deep Graph Library (DGL).
The dimension of our embedding is set at 256. 
We have set the learning rate at 0.003, with the margin in the loss function determined to be 100. 
To mitigate the risk of overfitting, we incorporate an early stopping mechanism, employing a patience parameter of 2.
% We run our experiment on AMD EPYC 7742 processor with NVIDIA A100 GPU. 

\noindent \textbf{Baseline.}
In our evaluation, we benchmark our proposed methodologies against conventional embedding techniques previously employed in previous security event detection frameworks~\cite{shin2020cybersecurity, le2017sonar, kristiansen2020cti}, i.e.,  \textit{TF-IDF} and \textit{Word2Vec}.
We also compare with transformer encoder-based text embedding methods, \textit{BERT}~\cite{bert}, , \textit{BERTweet}~\cite{bertweet} and \textit{SecureBERT}~\cite{securebert}.
\textit{BERTweet} and \textit{SecureBERT} are domain-adapted language models, which are further fine-tuned on domain-specific corpora from Twitter and cybersecurity texts, respectively.
Additionally, we include the \textit{Llama2} generative large language model (LLM), specifically the Llama-2-7b-chat-hf, in our comparison. 

We also extend our comparison to graph-based tweet embedding methods \textit{GCN}~\cite{gcn}, \textit{GATv2}~\cite{gatv2}, and \textit{GraphSAGE}~\cite{graphsage}, which are proposed for social event detection on Twitter.
GATv2~\cite{gatv2} is an advanced version of GAT~\cite{gat}, which addresses the limitations of the static attention mechanism. 

For the TF-IDF implementation, we utilized the scikit-learn Python package. 
For Word2Vec, we employed the trained model available through spaCy\footnote{https://spacy.io}. 
In the case of transformer-based text embedding methods, we used pre-trained models from Hugging Face, which produce embeddings of a fixed size of 768 dimensions. 
For encoding models, we use the \texttt{[CLS]} token embedding to represent sentence-level embeddings. 
Conversely, as the \texttt{[CLS]} token is not utilized for the generative model, Llama2, we compute the mean of all token embeddings (size of 4096 dimensions). 

Additionally, following previous social event detection methods~\cite{ren2022known, cao2021knowledge}, we initially constructed a graph utilizing general entity recognition methods as implemented in spaCy. 
Subsequently, variations of Graph Neural Networks (GNN) like GCN, GATv2, and GraphSAGE were applied to obtain the tweet embeddings. 
GATv2 is an advanced version of GAT, which fixes the static attention problem of GAT~\cite{gat}.
The embedding size for these methods was set at 256. 
To ensure a fair comparison, we adopted the same configuration parameters (including learning rate, loss margin, and early stopping strategy) as those used in our proposed method.

\input{tables/clustering_result}
\noindent \textbf{Evaluation metric.}
To evaluate the embedding efficiency, we use three commonly used cluster evaluation metrics: Normalized Mutual Information (NMI), Adjusted Mutual Information (AMI), and Adjusted Rand Index (ARI).

(1) \textit{Normalized Mutual Information (NMI):}
NMI is a normalization of the Mutual Information (MI) score to scale the results between 0 and 1, with 1 indicating perfect agreement between clusters.

(2) \textit{Adjusted Mutual Information (AMI):}
AMI is a chance-corrected variant of the MI metric that accounts for the expected MI. 
AMI ranges from 0 to 1, with 1 indicating perfect agreement.

(3) \textit{Adjusted Rand Index (ARI):} 
ARI, a chance-corrected Rand Index, measures cluster assignment similarity through pairwise comparisons. 
It ranges from -1 (disagreement) to 1 (perfect agreement), with 0 indicating random agreement.

\noindent \textbf{Results and discussions}.
As shown in Table~\ref{tab:clustering_result}, our security event attribution-centric tweet embedding
outperforms all other baselines across all three evaluation metrics for all three test datasets.
Despite the substantial temporal gaps between the training set and test set 2 and 3, our model consistently outperforms existing baselines. 
This can be attributed to the utilization of a graph-based approach that captures STIX object-sharing patterns through the tweet relation graph, making the generated representation resilient to topic changes.

The results also highlight the limitation of considering only text when embedding tweets for clustering.
Interestingly, TF-IDF shows a comparable result to PLMs. 
This is because our dataset only contains security event-related tweets, mitigating the issue of keyword ambiguity commonly associated with TF-IDF.
While domain-specific models, such as SecureBERT and BERTweet, are further fine-tuned on domains related to security and Twitter text, they sometimes show comparable or even lower performance than general-domain PLMs, as they are not optimized for clustering tweets for event detection.
Graph-based embedding methods, designed to enhance social event detection, prove less effective compared to our approach. 
This inefficiency stems from their inability to identify security-related entities in tweets, resulting in a graph-constructed lack of information to distinguish different security events. 
To summarize, the outstanding performance of \model{} can be attributed to its ability to integrate the tweet relation graph for capturing STIX object-sharing patterns between tweets and event-related features such as tweet content and temporal information. 
This integration ensures a more accurate and resilient security event detection, effectively adapting to topic changes in security events over time.

%% file: tables/clustering_dataset_exp.tex
\begin{table}[t!]
\caption{Training, validation, test data statistics.}
\ra{1.1}
\label{tab:clustering_dataset_exp}
\begin{adjustbox}{width=.9\columnwidth, center}
\begin{tabular}{lccc}
\toprule
& \textbf{\# Event} & \textbf{\# Tweet} & \textbf{Period} \\
\midrule
Training & 91 & 935 & 2022.06.01 $\sim$ 2022.09.01 \\
Validation & 23 & 328 & 2022.09.01 $\sim$ 2022.09.25 \\
\midrule
Test-1 & 24 & 353 & 2022.09.25 $\sim$ 2022.10.21 \\
Test-2 & 21 & 204 & 2024.01.01 $\sim$ 2024.01.31 \\
Test-3 & 23 & 234 & 2024.02.01 $\sim$ 2024.02.28 \\
\midrule
\textbf{Total} & 182 & 2,054 & \makecell[l]{2022.06.01 $\sim$ 2022.10.21 \\ 2024.01.01 $\sim$ 2024.02.28} \\
\bottomrule
\end{tabular}
\end{adjustbox}
\vspace{-5pt}
\end{table}

%% file: tables/clustering_result.tex
\begin{table*}[t]
\footnotesize
\ra{1.1}
\caption{Tweet Clustering Results. $\uparrow$: higher the better}
\label{tab:clustering_result}
\begin{adjustbox}{width=0.85\linewidth, center}
\begin{tabular}{clccc|ccc|ccc}
\toprule
& \multirow{2}{*}{\textbf{Model}} & \multicolumn{3}{c}{\textbf{Test-1}} & \multicolumn{3}{c}{\textbf{Test-2}} & \multicolumn{3}{c}{\textbf{Test-3}} \\
\cmidrule(lr){3-5}  \cmidrule(lr){6-8} \cmidrule(lr){9-11}
&  & \textbf{AMI} ($\uparrow$) & \textbf{ARI} ($\uparrow$) & \textbf{NMI} ($\uparrow$) & \textbf{AMI} ($\uparrow$) & \textbf{ARI} ($\uparrow$) & \textbf{NMI} ($\uparrow$) & \textbf{AMI} ($\uparrow$) & \textbf{ARI} ($\uparrow$) & \textbf{NMI} ($\uparrow$) \\
\midrule
\multirow{2}{*}{Keyword} & TF-IDF & 0.3036 & 0.0552 & 0.5147 & 0.4559 & 0.0989 & 0.6150 & 0.4669 & 0.0941 & 0.6218 \\
& Word2Vec & 0.0463 & 0.0060 & 0.1135 & 0.2469 & 0.0389 & 0.3876 & 0.1380 & 0.0246 & 0.2367 \\
\midrule
\multirow{4}{*}{\makecell[c]{PLM}} & BERT & 0.2389 & 0.0203 & 0.4395 & 0.2671 & 0.0291 & 0.4544 & 0.2716 & 0.0264 & 0.4573 \\
& BERTweet & 0.3466 & 0.0676 & 0.5211 & 0.2777 & 0.0312 & 0.4618 & 0.1729 & 0.0143 & 0.3372 \\
& SecureBERT & 0.3046 & 0.0324 & 0.5020 & 0.2555 & 0.0259 & 0.4434 & 0.1927 & 0.0161 & 0.3763 \\
& Llama2 & 0.1138 & 0.0090 & 0.3041 & 0.2324 & 0.0271 & 0.4375 & 0.2127 & 0.0254 & 0.4118 \\
\midrule
\multirow{3}{*}{Graph} & GCN & 0.2806 & 0.0869 & 0.4455 & 0.3771 & 0.1253 & 0.5454 & 0.3335 & 0.1171 & 0.5111 \\
& GATv2 & 0.3396 & 0.0988 & 0.5274 & 0.4065 & 0.1430 & 0.5476 & 0.3440 & 0.1112 & 0.5077 \\
& GraphSAGE & 0.3164 & 0.0912 & 0.5019 & 0.3666 & 0.1305 & 0.5350 & 0.3210 & 0.1048 & 0.4866 \\
\midrule
\multicolumn{2}{c}{\footnotesize \makecell[c]{Our Embedding}} & \textbf{0.5919} & \textbf{0.3384} & \textbf{0.7344} & \textbf{0.6561} & \textbf{0.4470} & \textbf{0.7763} & \textbf{0.5950} & \textbf{0.3387} & \textbf{0.7404} \\
\bottomrule
\end{tabular}
\end{adjustbox}
\vspace{-5pt}
\end{table*}

%% file: sections/4_framework.tex
\section{\model{}: Design and Implementation}
This section outlines the workflow for integrating the proposed security event attribution-centric tweet embedding method into a security event detection framework.
We begin by presenting the overall workflow of security event detection framework, followed by detailed discussions of two essential steps: the Tweet Category Tagging and the Event Identification, both of which are critical for the effective deployment of the proposed event-centric tweet embedding method in real-world settings.
\begin{figure}[t!]
    \centering
    \includegraphics[width=.8\columnwidth]{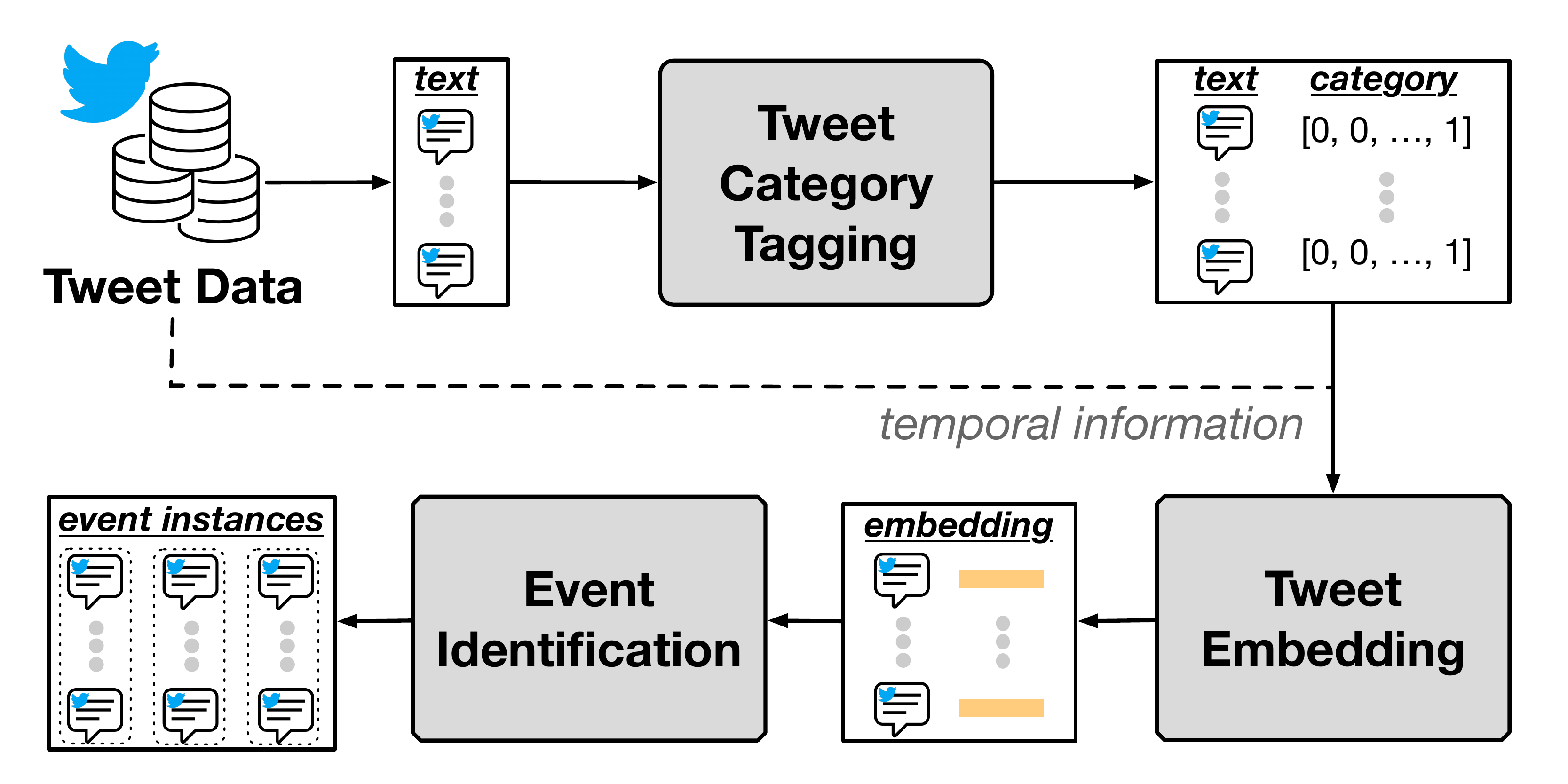}
    \caption{Overall workflow of \model{}.}
    \label{fig:overview}
    \vspace{-5pt}
\end{figure}
\subsection{Overall Workflow}
\label{sec:overall_workflow}
As shown in Figure~\ref{fig:overview}, we first retrieve tweets by using the Twitter Enterprise API\footnote{https://developer.twitter.com/en/docs/twitter-api/enterprise}.
Retrieving the entire tweets is impractical due to their extensive volume, so we opted to collect tweets based on a list of predefined security-related keywords which was aggregated by analysts from a cybersecurity company for the purpose of monitoring Twitter data.
The collected tweets are then fed into the Tweet Category Tagger to obtain category information. 
Tweets related to security are subsequently forwarded to the Tweet Embedding step, along with their temporal information. 
During this step, specialized embeddings are generated to enhance the detection of security events.
Subsequently, these embeddings undergo clustering and filtering in the Event Identification step to identify the security events.
As an output, we can get event instances, which are collections of tweets associated with individual security events.

\subsection{Tweet Category Tagging}
\label{sec:tweet_categorization}

While tweets are collected using predetermined security-related keywords, the collection inevitably includes some tweets that are not pertinent to security. 
Therefore, removing these irrelevant tweets in the initial phase significantly benefits the subsequent embedding phase by reducing additional noise.
Furthermore, categorizing events with specific types of information can aid security practitioners in analyzing and classifying these events. 
In this stage, we preprocess each tweet, ensuring the assignment of appropriate category tags.
Recognizing the multi-dimensional nature of security events, we have constructed our own multi-label tweet category dataset.
Based on this dataset, we design a multi-label classifier based on a Pretrained Language Model (PLM). 
\begin{figure}[t!]
    \centering
    \includegraphics[width=.9\columnwidth]{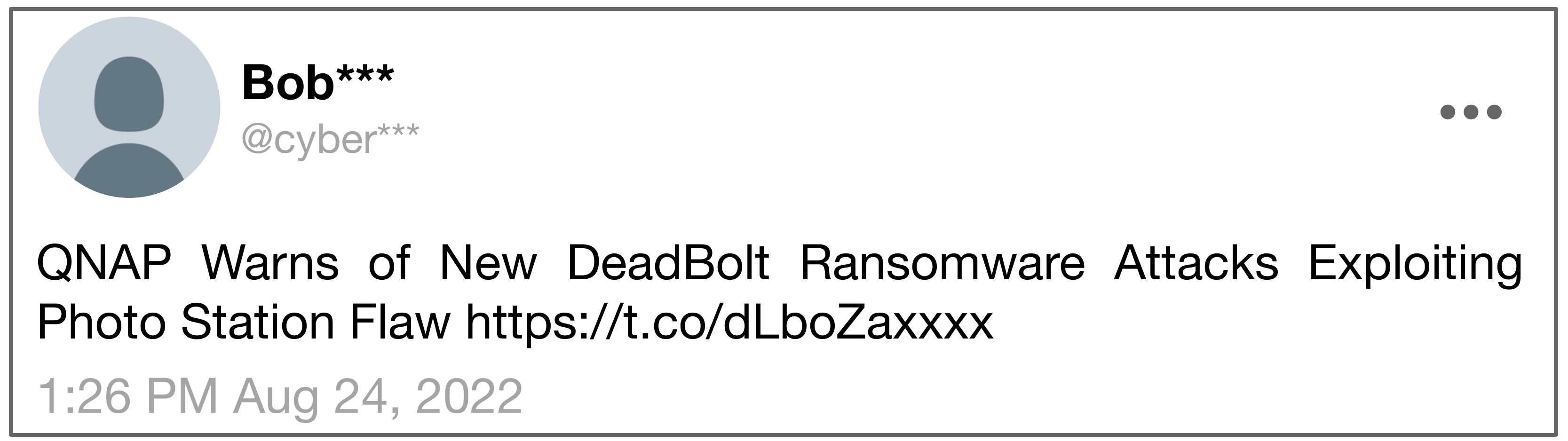}
    \caption{An example of a multi-category tweet: The above tweet pertains to both the \textit{Malware} and \textit{Vulnerability} categories.}
    \label{fig:motivating_example_multi}
    \vspace{-5pt}
\end{figure}

\noindent\textbf{Multi-label nature of security events.}
Previous research~\cite{ritter2015weakly, cydec, TwitterThreats} on event detection has automated categorization of tweets into predefined categories, such as \textit{DDoS}, \textit{Malware}, \textit{Vulnerability}, etc.
In practice, however, these categories are not mutually exclusive.
Real-world events often span multiple security categories, as exemplified by the tweet in Figure~\ref{fig:motivating_example_multi}.
The tweet reports on how a \textit{malware} (DeadBolt Ransomware) has been exploiting a \textit{vulnerability} (Photo Station Flaw) and therefore should be recognized as relevant to both categories.
Furthermore, when tweets pertain to multiple categories, it can introduce complications during the annotation phase. 
Suppose annotators are instructed to assign multi-label tweets into a single category. 
In that case, different annotators may assign similar tweets to different categories, thereby adversely affecting the performance of the classification.

\input{tables/data_category}
\noindent\textbf{Multi-label tweet dataset construction.}
To account for the multi-category nature of security tweets, we create a new dataset that labels all applicable security categories of each tweet.
We identify five cybersecurity categories and perform multi-label annotations for these categories.
Additionally, we have a separate category of "uninformative" tweets for tweets that are somewhat related to cybersecurity but lack sufficient information to be confidently categorized into specific event topics. 
To enhance annotation consistency and automatic classification, tweets outside of the five cybersecurity categories are distinguished between ``uninformative'' tweets and ``non-security'' tweets.
In total, our categorization framework encompasses seven categories, as presented in Table~\ref{tab:data_category}. 

Annotations were carried out by five security experts from a company specializing in cyber threat intelligence.
To assess the quality of the annotations, we measure inter-annotator agreement using the commonly used Fleiss' Kappa score. 
Our five annotators achieved an average Fleiss Kappa score of 0.75, indicating substantial agreement among the annotators according to the established guidelines~\cite{landis1977measurement}.
A total of 10,392 tweets were annotated, and their distributions can also be found in Table~\ref{tab:data_category}.
Among the tweets classified into specific categories, a total of 336 tweets (9.84\%) were multi-labeled, meaning they were associated with more than one category.
Among the multi-labeled tweets, 311 had two labels assigned, 26 had three labels assigned, and 1 had four labels assigned.

\begin{figure}[t!]
    \centering
    \includegraphics[width=.85\columnwidth]{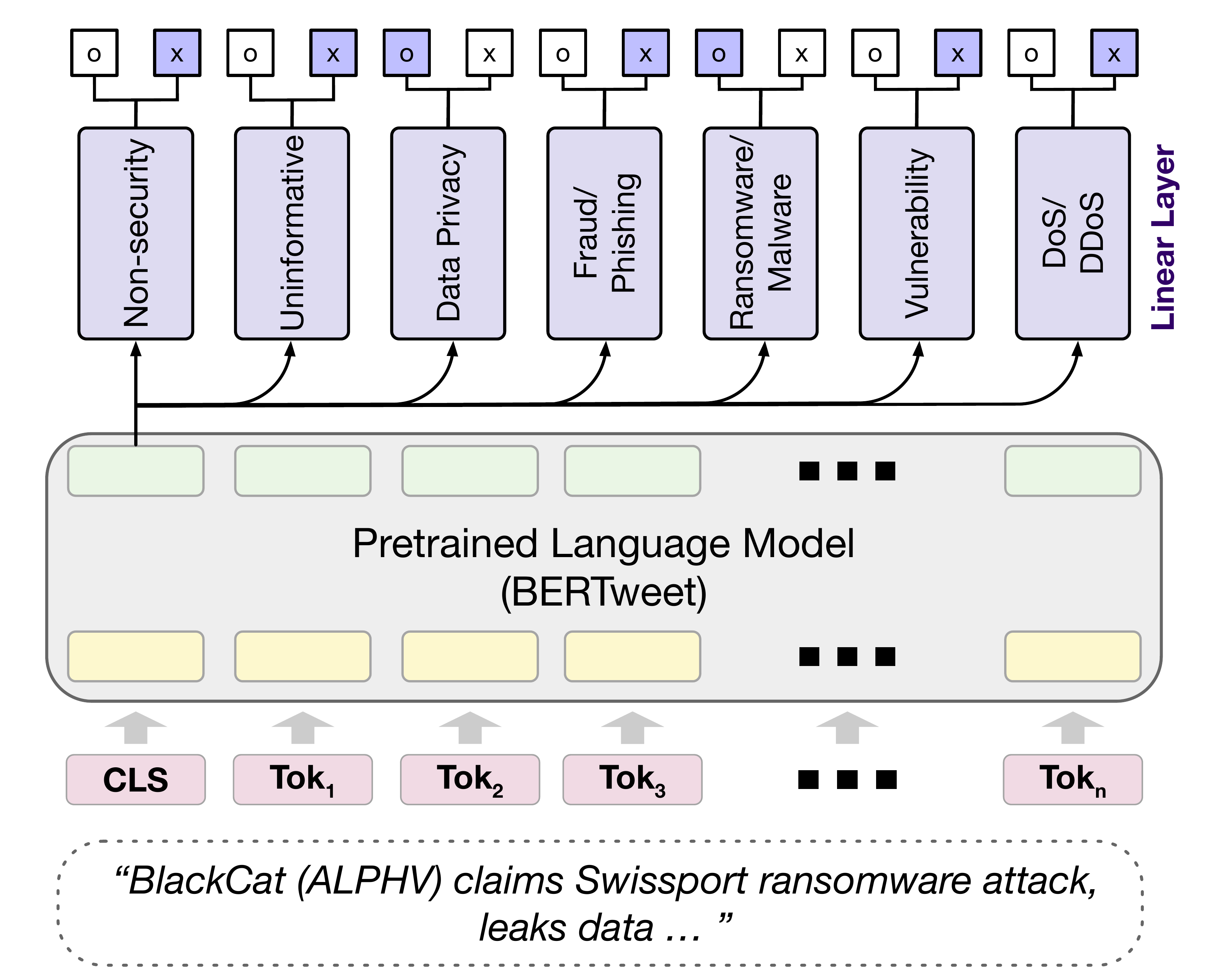}
    \caption{Architecture of Multi-label Tweet Categorizer}
    \label{fig:tweet_classifier}
    \vspace{-5pt}
\end{figure}

\noindent\textbf{Architecture and training.}
The Multi-label Tweet Categorizer is based on a Pretrained Language Model (PLM), which has shown great capability to capture complex contextual information and semantic relationships in text.
To achieve multi-category assignment, the classifier has a linear layer for each security category on top of the PLM output (embedding of \texttt{[CLS]} token).
With the dataset constructed, the weights of each linear layer and PLM weights are optimized with the Adam optimizer and cross-entropy loss.
The total loss is the sum of the cross-entropy losses calculated for each category. i.e., 
\begin{equation}
\mathcal{L}_c = \sum_{c \in \mathcal{C}} [-\frac{1}{N} \sum_{i}^{N}\mathbf{y}_i \cdot \log \hat{\mathbf{y}}_{i} + (1-\mathbf{y}_i) \cdot \log (1 - \hat{\mathbf{y}}_{i})]
\end{equation}
where $\mathcal{C}$ is the set of security categories, $N$ is the number of samples in each batch, $\mathbf{y}_i$ is the true label of $i$-th sample, and $\hat{\mathbf{y}}_{i}$ is the predicted label of $i$-th sample.
A loss function is calculated separately for each category and then accumulated to optimize the classifier.
The classifier was trained using a batch size of 64 and a learning rate of 1e-5, and an early stopping strategy was adopted with a patience of 5 to prevent overfitting.
A weighted cross entropy loss (with a weight of 0.8 on the positive class) is utilized to tackle the class imbalance issue in the training data.
The trained Multi-label Tweet Categorizer achieved a 0.8322 F1-score on average.
The detailed experimental setup and result are shown in Appendix~\ref{sec:multilable_performance}.

\input{tables/eval3_precision_recall}
\subsection{Event Identification}
The embeddings generated by our method are clustered into event instances within this component, and clusters lacking sufficient information are filtered out.
The DBSCAN~\cite{DBSCAN} clustering algorithm is employed for this purpose. 
DBSCAN, which stands for Density-Based Spatial Clustering of Applications with Noise, is a widely used algorithm for unsupervised density-based clustering.
A key advantage of DBSCAN is its ability to determine clusters without requiring a number of clusters to be defined beforehand.
This makes it an appropriate choice for event detection, as the exact number of events is not possible to be determined in advance in the wild.

After clustering, the collection of tweets will be organized into clusters, each representing specific events.
Using a heuristic measure of relevance, clusters are filtered to keep only high-relevance events.
The clusters are filtered according to the following formula.
\begin{equation}
\label{eq:clsuter_filtering}
Score(C) = \frac{\#\ \ User}{\#\ \ Tweet}    
\end{equation}
Clusters with a score lower than a threshold value are removed, as such events include those repetitively posted by a few users or by a single automated account.
These accounts often post repetitive content within a short time frame, resulting in an abundance of tweets that share similar entities and features.
Consequently, these tweets become interconnected in the tweet relation graph, ultimately producing similar embeddings for these spamming tweets and leading to the formation of clusters that are not event-related.
However, these clusters can be easily filtered out due to the relatively low engagement from users compared to clusters that are actually related to events. 
Using cumulative distribution of scores calculated for clusters extracted from \model{}, we found that over 60\% of clusters have scores exceeding 0.80\footnote{the result can be found in our released artifact}. 
Consequently, in our study, clusters with scores below the threshold of 0.80 are excluded from further analysis.

%% file: tables/data_category.tex
\begin{table*}[t!]
\ra{1.1}
\caption{Statistics of categories in the tweet classification dataset along with annotation guidelines.}
\label{tab:data_category}
\footnotesize
\adjustbox{width=.95\linewidth, center}{
\begin{tabular}{ll rll}
\toprule
\multicolumn{2}{l}{\textbf{Category}}&\textbf{\# Tweets (Ratio)} &\textbf{Short guideline description} & \textbf{Example Tweet}\\
\midrule

\multicolumn{2}{l}{Non-security} & 5,286 (50.87\%) & \makecell[l]{Tweets not related to security, such as \\ daily tweets, job postings, etc,.} & \makecell[l]{ \textit{``Hiring senior cyber security engineer and penetration} \\ \textit{tester on remote roles''}} \\

\multicolumn{2}{l}{Uninformative} & 1,671 (16.08\%)  & \makecell[l]{Tweets that have limited information and thus \\ cannot be specified into a specific category} & \makecell[l]{\textit{``76\% of organizations worldwide expect to suffer} \\ \textit{ cyberattack this year''}} \\

\multicolumn{2}{l}{Security}& 3,435 (33.05\%) & & \\

& Vulnerability & 1,617 (15.56\%) & \makecell[l]{Tweets on newly discovered, exploited, \\ or patched vulnerabilities } & \makecell[l]{\textit{``... multiple vulnerabilities in the API and web-based} \\ \textit{management interfaces of Cisco ...''}} \\

& Ransomware/Malware & 742 (7.14\%) & \makecell[l]{Tweets on emerging malware/ransomware attacks,\\ updates in techniques or tactics, etc.} & \makecell[l]{\textit{``... by targeting U.S. organizations with \#ransomware} \\ \textit{attacks. \#CobaltMirage''}} \\

& Data Privacy & 722 (6.95\%) & \makecell[l]{Tweets on data breaches, information disclosure, \\ account hijacking, eavesdropping, etc.} & \makecell[l]{\textit{``Axonius SaaS Management identifies misconfigurations} \\ \textit{and data security risks''}} \\

& Fraud/Phishing & 348 (3.35\%)  & \makecell[l]{Tweets on fraud/phishing techniques or campaigns,  \\ such as email phishing, SMS scam campaigns, etc.} & \makecell[l]{\textit{``Microsoft: hackers are using open source software} \\ \textit{and fake jobs in phishing attacks''}} \\

& DoS/DDoS &  336 (3.23\%)  & \makecell[l]{Tweets on Dos/DDoS campaigns} & \makecell[l]{\textit{``Cloudflare says it thwarted record-breaking HTTPS} \\ \textit{DDoS flood''}} \\

\cline{2-3}
& Multi-labeled & 338 (9.84\%) & \\
\cline{1-3}
\multicolumn{2}{l}{\textbf{Total}} & \textbf{10,392 (100.0\%)} &\\
\bottomrule
\end{tabular}
}
\end{table*}

%% file: tables/eval3_precision_recall.tex
\begin{table*}[t!]
\footnotesize
\ra{1.1}
\caption{Event detection performance. Note that precision measures the percentage of correctly identified events among the extracted events, and recall measures the percentage of extracted events compared to the events published by news publishers.}
\label{tab:eval_wild}
\begin{adjustbox}{width=.9\linewidth, center}
\begin{tabular}{crllllll}
\toprule
&  & Data Privacy & Fraud/Phishing & Ransomware/Malware & DoS/DDoS & Vulnerability & Total\\
\midrule
\multirow{3}{*}{Precision} & SONAR & 0.2593 (14/54) & 0.4000 (6/15) & 0.3924 (31/79) & 0.5000 (4/8) &0.5238 (11/21) & 0.3522 (56/159)\\
& W2E & \textbf{0.6667 (2/3)} & \textbf{1.0000} (3/3) & \textbf{0.7500 (3/4)} & 0.0000 (0/0) & \textbf{1.0000} (2/2) & \textbf{0.8182 (9/11)} \\
& \model{} & 0.4766 (61/128) & 0.5490 (28/51) & 0.7282 (142/195) & \textbf{0.5625 (9/16)} & 0.6912 (94/136) & 0.6415 (272/424) \\
\midrule
\multirow{3}{*}{Recall} & SONAR &0.3824 (13/34) & 0.2727 (6/22) & 0.3239 (23/71) & 0.6667 (1/3) & 0.2045 (9/44) & 0.2980 (45/151) \\
& W2E & 0.0000 (0/34) & 0.0000 (0/22) & 0.0282 (2/71) & 0.0000 (0/3) & 0.0455 (2/44) & 0.0265 (4/151)\\
& \model{} & \textbf{0.4706 (16/34)} & \textbf{0.4091 (9/22)} & \textbf{0.5915 (42/71)} & \textbf{0.6667 (1/3)} &  \textbf{0.5909 (26/44)} &  \textbf{0.5563} (84/151) \\
\bottomrule
\end{tabular}
\end{adjustbox}
% \vspace{-5pt}
\end{table*}

%% file: sections/4_evaluation.tex
\subsection{Evaluation}

In the evaluation, our main focus is to address the following three questions.
\begin{itemize}
    \item How effective is our framework in identifying events in real-world scenarios compared to the existing security event detection framework? 
    \item How does each component of our framework (features or GNN models) affect the performance of clustering effectiveness? 
\end{itemize}

\subsubsection{Event Detection Performance}
\label{sec:exp_wild}

In this section, we demonstrate the practical applicability of the complete framework in real-world scenarios.
The detection result is compared with the existing security event detection framework.

\noindent \textbf{Dataset.}
The study is conducted using a dataset consisting of tweets collected in November 2022.
As described in Section~\ref{sec:overall_workflow}, tweets are acquired through Twitter Enterprise API with predefined security-related keywords employed by cybersecurity analysts for monitoring purposes. 
A total of 5,291,166 tweets from 1,221,332 distinct users are collected in this period. 

\noindent \textbf{Experimental setup.}
To evaluate the effectiveness of our framework, we extracted events using \model{} and existing event detection frameworks~\cite{le2017sonar, shin2020cybersecurity} and compared their precision and recall.
\emph{Precision} measures the percentage of correctly identified events among the extracted events, while \emph{recall} measures the percentage of retrieved events compared to the total events published by established security news publishers. 
To calculate recall, we curate a list of security events by examining articles published by BleepingComputer, The Hacker News, and Hackmageddon throughout November 2022. 
From these sources, we obtained a total of 167 unique events.
To ensure accuracy, two security experts were involved in the assessment.

\noindent \textbf{Results and discussion.}
The evaluation results are shown in Table~\ref{tab:eval_wild}. 
In terms of both precision and recall, our methods outperform existing security event detection frameworks in most cases.
Notably, our framework detects a significantly greater number of events compared to the baseline methods.

The overall precision of \model{} is 0.6415, indicating that the majority of the events identified by our framework were indeed genuine events.
Although W2E shows higher precision than \model{}, it only detected 9 security events, which is much smaller than the number of events detected by \model{}.
However, there are still some clusters erroneously identified as events by our framework if tweets discussed security topics that were not events.
For example, a cluster may consist of multiple tweets sharing a blog post titled \textit{``Ransomware, storage and backup: Impacts, limits, and capabilities''}.
Even though this cluster does not represent a security event, the discussion was related to security and was shared by multiple people.
This demonstrates a limitation that not all influential security-related discussions are indicative of security events.
Nevertheless, the high precision of \model{} suggests that such cases are rare compared to actual event instances.

Among the 151 events reported in the examined news outlets, \model{} successfully identified a total of 84 events, and the other methods SONAR and W2E identified 45 and 4 events, respectively.
However, it is important to note that there were cases where events were not captured by any method.
For instance, some events may have a limited number of associated tweets, as we are using a fixed list of keywords to collect tweets.
This limitation is inevitable since collecting all tweets on the platform is impractical due to the sheer volume of data and associated costs (Twitter API charges per tweet).
Thus, events that do not have overlapping keywords with our predefined list used for collecting tweets may have only a small number of tweets collected. 
As a result, these events may fail to form distinct clusters or could be excluded during the clustering phase during filtering.
For instance, consider the event \textit{``Dell, HP, and Lenovo Devices Found Using Outdated OpenSSL Versions''}, which had only 5 tweets available in our tweets collection.
Such an event would be filtered out as there were not enough tweets to support the cluster.

\subsubsection{Ablation Study}
In this section, we assess the performance differences of our framework caused by variations in the initial features and the GNN model used for event-centric embeddings.
The selection and representation of different features can affect the output event-centric embedding.
Similarly, the choice of the GNN model for generating event-centric embeddings can also influence the performance of the framework. 
Different GNN architectures, such as Graph Convolutional Networks (GCNs)~\cite{gcn}, GraphSAGE~\cite{graphsage}, or Graph Attention Networks (GATs)~\cite{gatv2}, can be employed to capture the relational information among tweets and generate event embeddings. 
Each GNN model has its own strengths and weaknesses, and selecting the most suitable model can improve the quality of the event embeddings and the overall performance of the framework.

To assess the performance differences caused by these modules, various experiments can be conducted. 
For example, different combinations of initial features can be evaluated to identify the most effective feature set for event detection and classification. 
Different text embedding methods can also be evaluated to identify the most effective text embedding methods in our framework. 
Additionally, multiple GNN models can be compared in terms of their ability to capture event information and generate informative event attribution-centric tweet embeddings.

\input{tables/ablation_all}

\noindent \textbf{Results and discussion.}
Table~\ref{tab:ablation_all} presents the performance outcomes obtained by manipulating the initial features mentioned in Section~\ref{sec:embedding_generator}.
\model{} incorporating all features exhibits the highest performance, indicating the significance of the event-related features. 
The absence of tweet content results in the poorest performance, highlighting the importance of tweet content within our framework.
Without temporal feature, the clustering performance degrades from 0.7344 to 0.6968 in terms of NMI score; and without text feature, the clustering performance drops from 0.7344 to 0.6519.

Regarding text features, as demonstrated in Table~\ref{tab:ablation_all}, the utilization of BERTweet for embedding tweet content yields the highest performance.
This can be attributed to the specific pretraining of BERTweet on tweet text, which allows it to capture the semantics and characteristics of tweets effectively.
Conversely, despite being trained in security-related contexts, SecureBERT fails to demonstrate satisfactory performance due to its limited adaptation to tweet text. 
Furthermore, Table~\ref{tab:ablation_all} reveals that GATv2 outperforms other graph neural network (GNN) models in terms of performance, which can be attributed to its attention mechanism that effectively mitigates the impact of false connections in the graph. 
Based on these evaluation findings, our \model{} selects BERTweet for tweet content embedding and GATv2 for generating event attribute-centric embeddings.

\noindent \textbf{Importance of Temporal feature. }
Table~\ref{tab:ablation_all} demonstrates that incorporating temporal features leads to improved clustering performance, resulting in more accurate event identification. 
We present cases where event identification benefits significantly from utilizing temporal features.

Without temporal features, we observed that some tweets that should belong to the same cluster are identified as separate events. 
For instance, we found that the tweets related to the Juniper Junos OS vulnerability disclosure is actually separated into two different clusters: one discussing the event itself, e.g., \textit{High-Severity Flaws in Juniper Junos OS Affect Enterprise Networking Devices''}, and another providing detailed security implications, e.g., \textit{…Multiple high-severity … some of which could be exploited to achieve code execution. Chief among them is…’’}. 
However, when temporal features were included, these tweets were accurately clustered into a single event.

Additionally, without temporal features, tweets pertaining to different events were erroneously clustered together. 
A case observed in our further analysis found that four tweets are clustered into the same event without temporal feature.
These tweets actually described two distinct events: the “Prestige” ransomware impacting organizations in Ukraine and Poland, and the Chinese ‘Spyder Loader’ Malware targeting organizations in Hong Kong. 
The high semantic similarity and lack of explicit mentions of target organizations caused their embeddings to be similar.
However, with temporal features, these tweets were correctly clustered into their respective events.

These cases substantiate that temporal features are crucial for enhancing security event detection performance. 
Moreover, the number of tweets classified as noise by the DBSCAN clustering algorithm decreased from 98 to 43 when temporal features were included, further emphasizing its importance.

\subsubsection{Response time}
\input{tables/response_time}
In order to extract security events in a timely manner, \model{} needs to process a large volume of events efficiently. 
This section evaluates the response time of \model{}.
Following Figure~\ref{fig:overview}, we break down the response time for each step, including Tweet Category Tagging, Tweet Embedding, and Event Identification.
For the Tweet Embedding process, we also present the response times for two important subprocesses: STIX object detection and graph embedding generation as explained in Section~\ref{sec:emb_method}. 

\noindent \textbf{Experimental Setup.}
We use 44,093 tweets, which is the average number in a 6-hour window for the tweets collected in November 2022.
6-hour time window, is also used for analyzing the security event trends in the following Section~\ref{sec:measurement_casestudy}-A. 
The experiment was conducted on a Tesla V100 GPU and an Intel Xeon CPU.

\noindent \textbf{Results \& Analysis.}
As shown in Table~\ref{tab:process_times}, the categorization of 44,093 tweets took 96.3 seconds. 
Tweet embedding required a total of 2100.4 seconds, with STIX object detection taking 1966.2 seconds and embedding generation taking 34.2 seconds. 
The final event identification took 0.25 seconds. 
This time represents an upper bound, as the entire process can be pipelined: while processing these tweets, newly arrived tweets can undergo categorization and embedding generation concurrently, allowing this information to be utilized in the next processing cycle.
The results indicate that \model{} is capable of processing a substantial volume of tweets within a reasonable timeframe. 

%% file: tables/ablation_all.tex
\begin{table}[t!]
\caption{GNN model ablation study results. $\uparrow$: The higher the better. $\mathbf{\neg}$ indicates the model without the following features. The boldface represents the best performance.}
\label{tab:ablation_all}
\ra{1.1}
\begin{adjustbox}{width=.9\columnwidth, center}
\begin{tabular}{clccc}
\toprule
& & \textbf{AMI} ($\uparrow$) & \textbf{ARI} ($\uparrow$)& \textbf{NMI} ($\uparrow$)\\
\midrule
\multirow{3}{*}{\makecell[c]{Node \\ Feature}} & \makecell[l]{{\footnotesize w/o \texttt{tweet}}} & 0.5117 & 0.2100 & 0.6519 \\
& \makecell[l]{{\footnotesize w/o \texttt{temporal}}} & 0.5629 & 0.2136 & 0.6968 \\
& \makecell[l]{{\footnotesize \texttt{all}}} &  \textbf{0.5919} & \textbf{0.3384} & \textbf{0.7344}\\
\midrule
\multirow{3}{*}{\makecell[c]{Text \\ Feature}} & \makecell[l]{\model { \footnotesize(\texttt{BERT})}} & 0.5407 & 0.2397 & 0.6631 \\
& \makecell[l]{\model { \footnotesize(\texttt{SecureBERT})}}  & 0.5208 & 0.2378 & 0.6464 \\
& \makecell[l]{\model { \footnotesize(\texttt{BERTweet})}} &  \textbf{0.5919} & \textbf{0.3384} & \textbf{0.7344}\\
\midrule
\multirow{3}{*}{\makecell[c]{GNN \\ Model}} & \makecell[l]{\model { \footnotesize(\texttt{GCN})}} & 0.2148 & 0.0502 & 0.4454 \\
& \makecell[l]{\model { \footnotesize(\texttt{GraphSAGE})}}  & 0.5002 & 0.2013 & 0.6535 \\
& \makecell[l]{\model { \footnotesize(\texttt{GATv2})}} & \textbf{0.5919} & \textbf{0.3384} & \textbf{0.7344}\\
\bottomrule
\end{tabular}
\end{adjustbox}
\vspace{-5pt}
\end{table}

%% file: tables/response_time.tex
\begin{table}[t!]
\centering
\ra{1.1}
\caption{Response time of \model{}}
\label{tab:process_times}
\begin{adjustbox}{width=.65\columnwidth, center}
\begin{tabular}{lrr}
\toprule
\textbf{Process}                                       && \textbf{Time (s)} \\ 
\midrule
\textbf{Tweet Category Tagging}                        && 96.3              \\
\textbf{Tweet Embedding (Total)}                       && 2100.4            \\
\quad \textit{- STIX Object Detection}                 && 1966.2            \\
\quad \textit{- Embedding Generation}                  && 34.2              \\
\textbf{Event Identification}                          && 0.25              \\ 
\bottomrule
\end{tabular}
\end{adjustbox}
\vspace{-5pt}
\end{table}

%% file: sections/5_usecases.tex
\section{Measurement and Case Studies}
\label{sec:measurement_casestudy}
\begin{figure*}[t!]
    \centering
    \includegraphics[width=.9\textwidth]{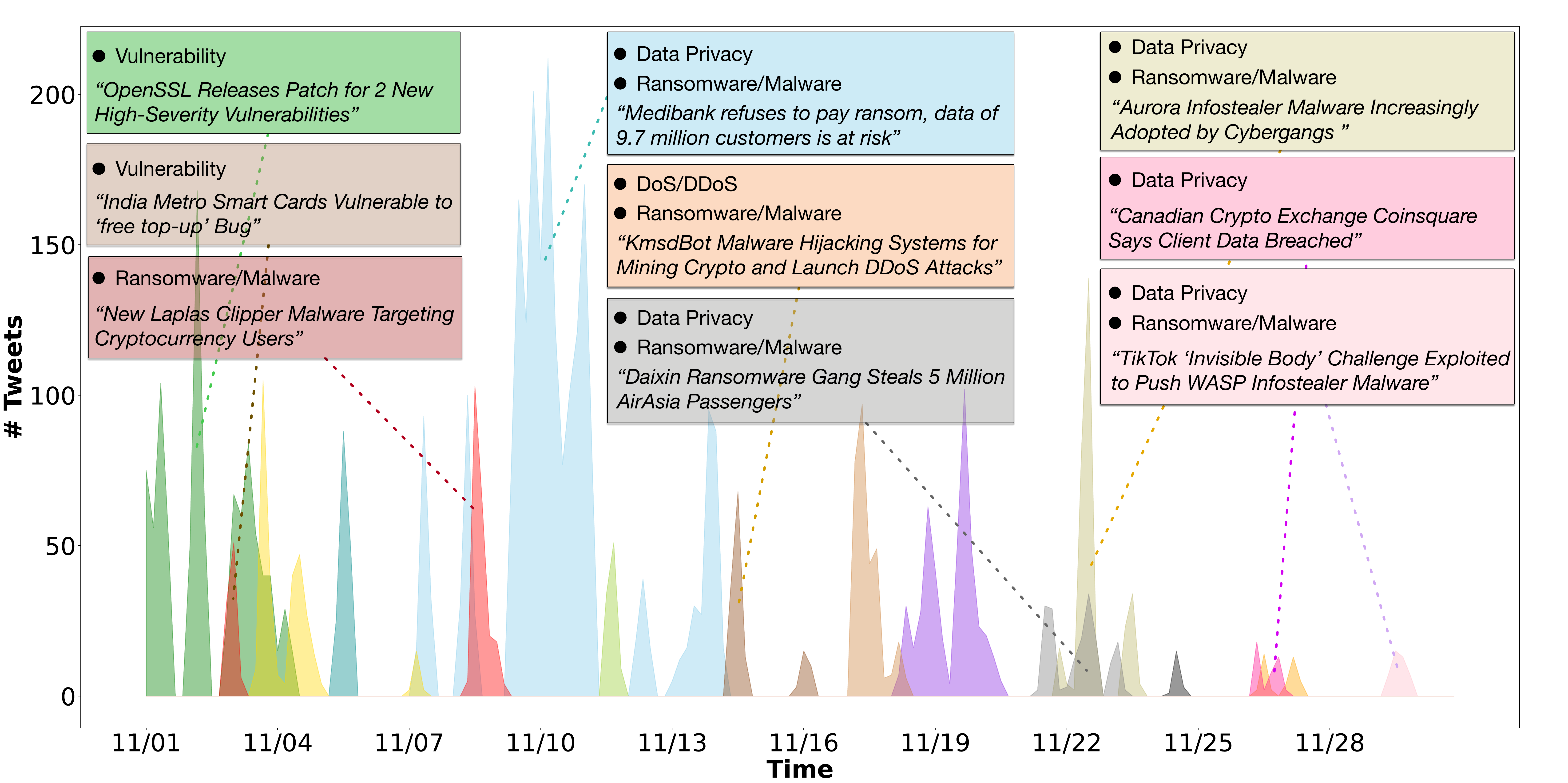}
    \caption{Visualization of events detected by \model{} between November 1 and November 30, 2022, with distinct colors denoting different events. The extent of each colored region correlates with the volume of tweets related to its event.}
    \label{fig:usecase_trend}
\end{figure*}
We here demonstrate two use cases of our framework: analyzing security event trends and finding informative security users.
Using our framework, we extract events between November 1, 2022, and November 30, 2022.
For every 4-hour period, events are extracted from tweets collected in the previous 6-hour period. 
In other words, events are extracted using a window size of 6 hours and a stride of 4 hours.
It's important to note that these parameters can be adjusted as per specific requirements or preferences.

\subsection{Security Event Trend Analysis}

\model{}'s ability to detect security events allows for an empirical analysis of the threat landscape.
Monitoring event tweets can produce event insights that may not be easily deducible from traditional news outlets, such as event magnitude and event impact over time.
To demonstrate such insights, we present a compilation of real-world events detected by \model{}.
We perform security event trend analysis by using our framework to detect events that occurred between November 1, 2022, and November 30, 2022. 
The detected events in this timeframe and their associated tweets are depicted in Figure~\ref{fig:usecase_trend}. 

As can be seen in Figure~\ref{fig:usecase_trend}, a considerable number of tweets discuss the OpenSSL vulnerability patch event at the beginning of the month. 
On November 1st, the OpenSSL team issued an advisory that warned of two high-severity vulnerabilities (CVE-2022-3602 and CVE-2022-3786) affecting OpenSSL versions 3.0.0-3.0.6.
Given the popularity of OpenSSL and the severity of this vulnerability, this event elicited significant concern within the open-source community and beyond. 
In contrast, another critical vulnerability event, the ConnectWise Server RCE Vulnerability, was detected in a similar timeframe but garnered significantly less attention.
Although this particular vulnerability was also described as ``critical", it was limited only to the ConnectWise service management software platform and therefore had a limited impact compared to the OpenSSL vulnerability.
The discrepancy in volume between these two events reflects their respective levels of prominence, which can serve as an important indicator when determining the significance of a vulnerability. 
Considering the constrained availability of maintenance resources, security practitioners are required to prioritize the patching of vulnerabilities based on their severity levels. 
In this context, the volume of tweet samples can provide valuable information to aid security practitioners in assessing the importance of specific events.

% Advantage 2:  
Of all the events in Figure~\ref{fig:usecase_trend}, the Medibank data leak event had the highest volume of tweets and showed a prolonged impact. 
Medibank, a prominent player in the Australian private health insurance industry, suffered a data breach in October that allowed hackers to gain access to the personal data of 9.7 million customers.
This data breach event is considered as one of the biggest data breach events of the region~\cite{top10_databreach_1, top10_databreach_2}.
Tweets extracted from this timeframe depict a comprehensive picture of the data leak.
On November 7th, the event was first discussed on Twitter after Medibank released a statement refusing to pay the ransom demand made by the attackers.
On November 9th, the event received massive attention on Twitter after attackers released an initial batch of stolen data on the dark web.
Due to the massive scale and impact of these leaks, the event remained relevant on Twitter for days.
During this time, the talking points of tweets underwent constant evolution as Australian authorities continued efforts to mitigate damage while attackers continued to release data.
Although news outlets, such as BleepingComputer and The Hacker News, did cover the event itself, they failed to capture the ongoing developments that shaped the event.
On the other hand, the event's entire sequence of developments was observable through our framework, suggesting it can provide more detailed and comprehensive information compared to traditional news outlets.

Through \model{}'s detection results, it is possible to infer the importance of specific categories or topics.
The multi-label categorization capability of our framework can be used to identify important or trending topics during a given period.
Based on our detection results in November, we found a significant number of events that were classified as both the \textit{Data Privacy} and \textit{Ransomware/Malware} categories, accounting for 11.33\% of the total events. 
This suggests that a notable number of malware and ransomware operations were specifically threatening users' personal data during this period.
Some noteworthy examples within this category include the AirAsia data leak caused by the Daixin ransomware, WASP Infostealer malware spread via TikTok's `invisible body challenge', and the widespread adoption of the Aurora Infostealer. 

Another observation during this period is the large number of incidents specifically targeting crypto assets. 
On November 8th, cryptocurrency users became victims of a new malware called the Laplas Clipper. 
This Infostealer operates by monitoring a victim's clipboard activity and replacing any wallet addresses with the attacker's addresses. 
Around November 14th, the KmsdBot Malware was identified. 
This malware mines cryptocurrency and carries out DDoS attacks. 
It leverages the Secure Shell (SSH) cryptographic protocol to infiltrate targeted systems. 
On November 26th, a data breach targeting cryptocurrency users occurred.
Coinsquare, a major cryptocurrency exchange in Canada, publicly acknowledged that they had experienced data breaches affecting their customers.

\begin{figure}[t!]
    \centering
    \includegraphics[width=\columnwidth]{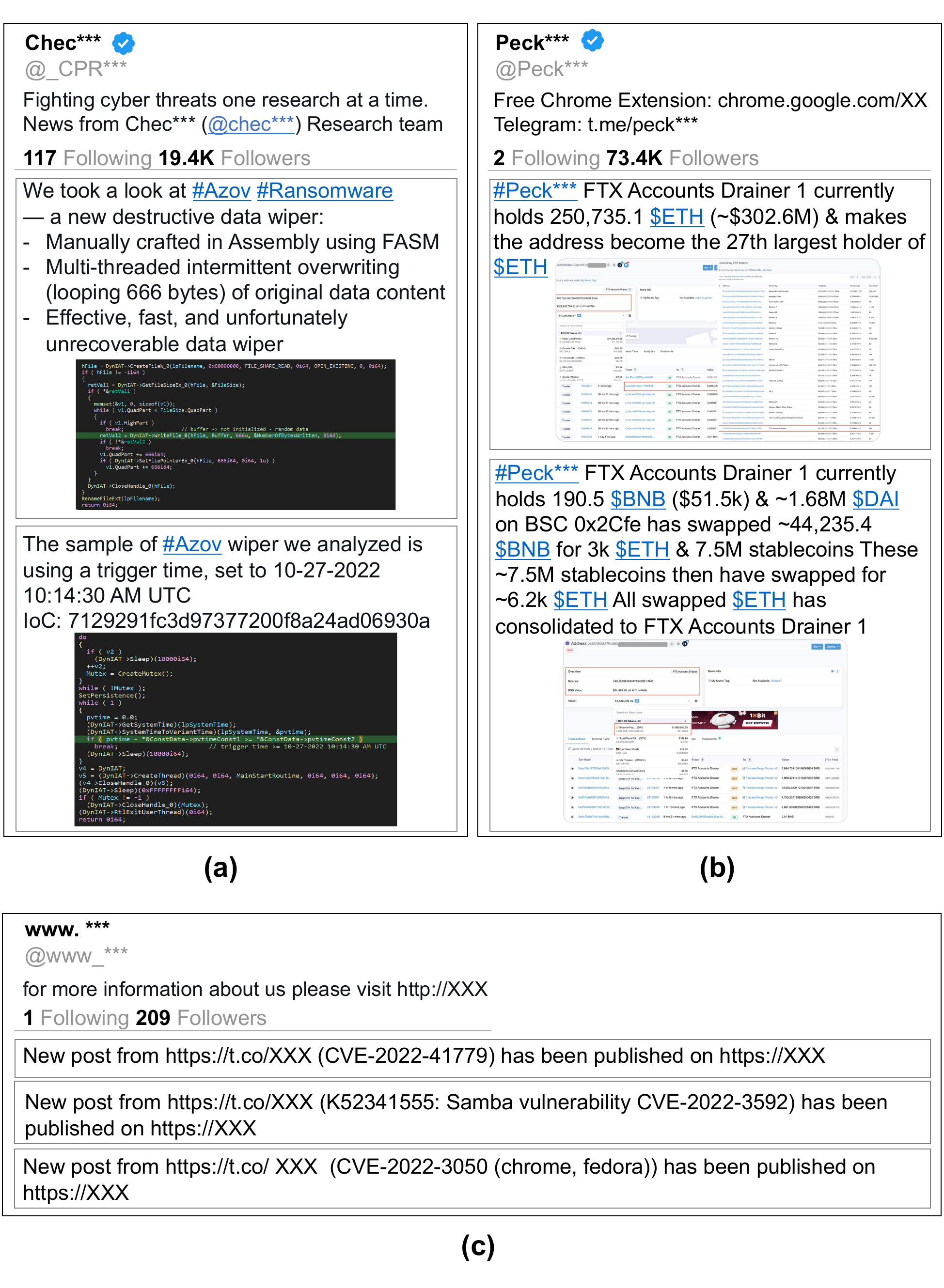}
    \caption{A comparison of top-ranked informative users extracted from two distinct methods. (a) and (b) shows the users extracted through our approach, while (c) shows the user extracted by counting the number of security-related tweets.}
    \label{fig:usecase_user_examples}
    \vspace{-7pt}
\end{figure}

\subsection{Finding Informative Security Users}

One extended application of \model{} is to find such informative security users.
On Twitter, there are informative security users who provide valuable and in-depth threat intelligence from events.
Informative users often go beyond simple news dissemination, providing additional valuable insights into security incidents.
It is valuable to identify such users since they often provide analysis on events before they come to public attention.

To find informative users, we take into account the following two key characteristics:
\begin{enumerate}
    \item High density of security event-related tweets.
    \item Have a substantial number of followers.
\end{enumerate}

This leverages the tendency of informative users to tweet often on security-related topics, as well as the tendency of these to have many followers, since their tweets are considered as reliable sources.

To extract users with the characteristics above, we use the following scoring function. 
For a user $U$, the score is formulated as: 
\begin{equation}
Score(U) = \frac{\sum_{t \in T} \frac{e_t}{N_t}} {{\vert T \vert }} \cdot log(F_u + 1)
\end{equation}
where $T$ is the set of time period (windows), $N_t$ is number of tweets tweeted by user $U$ in period $t$, $e_t$ is the number of event-related tweet tweeted by user $U$ in period $t$, and $F_u$ is the number of followers of user $U$.

$\sum_{t \in T} \frac{e_t}{N_t}$ measures the ``density'' of event-related tweets made by a user in period $t$, which rewards users with the first characteristic. 
$log(F_u + 1)$ accounts for the second characteristic by scaling the number of followers logarithmically.
This ensures that users with more followers rank higher in our scoring function.
Since our framework considers a number of categories separately, we can calculate the scores for each category. 

The samples of top-ranked users extracted with our approach is included in our released artifact. 
In addition, we show tweets of identified users in Figure~\ref{fig:usecase_user_examples}. 
One of the informative users, \textit{@\_CPR***}, represents a security research team.
This user frequently disseminates valuable security research results on malware and ransomware. 
This user elucidates malicious components in installers and shares Indicators of Compromise (IOCs). 
For example, Figure~\ref{fig:usecase_user_examples} (a) presents posts on the Azov wiper.
Azov was initially reported by others as ransomware, but the user corrected its classification to a wiper designed for maximum data destruction. 
Unlike other users that just mentioned the discovery of the wiper, \textit{@\_CPR***} shared research findings that further highlighted its dangers.

Figure~\ref{fig:usecase_user_examples} (b) illustrates the tweets from another identified informative user, \textit{@Peck***}. 
This user specializes in scrutinizing fraud and phishing events within the cryptocurrency realm. 
On November 11th, FTX, a large cryptocurrency exchange, suffered a hacking incident resulting in damages of over \$600 million.
\textit{@Peck***} conducted a comprehensive analysis of the cryptocurrency flows associated with the FTX hacking event, and shared results on Twitter.
Following this users would have allowed for faster information compared to relying on news outlets.

Identifying informative users is challenging without utilizing our framework.
As a baseline, we attempted to find informative users using solely the count of security-related tweets. 
Figure~\ref{fig:usecase_user_examples} (c) illustrates one of the top users identified using this count-based approach.
As shown in the figure, the user's tweets contain limited information.
The tweets are automated messages alerting whenever a new post is made on a site that lists Common Vulnerabilities and Exposures (CVEs).
While the tweets are linked to security-related information, the tweets themselves have limited information.
Consequently, these users cannot be considered informative users.
We find that users with a high volume of security tweets are not necessarily informative users.

%% file: sections/6_discussion.tex
\section{Discussion}
\label{sec:discussion}

\subsection{Adaptability of \model{} in other platform}

Twitter serves as a primary platform for information dissemination, motivating us to use it as the target platform for security event detection. 
Nevertheless, \model{} can also extend its applicability to other platforms. 
We validate our approach on Mastodon, an open-source, decentralized social media platform positioned as an alternative to centralized platforms like Twitter. 
We collect toots (counterparts to tweets) with a list of security-related hashtags in November 2022 through the Mastodon API\footnote{https://docs.joinmastodon.org/client/intro/}.
A total of 17,680 toots are collected in this period.

Despite being trained on Twitter data, the method proves effective in event detection on Mastodon.
We have identified a total of 257 events from Mastodon, most of which were also detected from Twitter.
Out of the 257 events detected on Mastodon, 156 were confirmed as true security incidents, but about 100 events were not detected on this platform.
Noteworthy events on Mastodon, such as the Medibank data leak and OpenSSL patch release, often aligned with Twitter detection results. 
However, certain major events, like the Coinsquare data breach and pro-Russian hacktivists' DDoS attack on the EU Parliament site, remained undetected. 
This can be attributed to Mastodon's smaller user base and the consequently fewer number of toots compared to Twitter. 
Additionally, minor events such as a new phishing campaign by Ducktail hackers and a phishing attack impersonating BAYC Founder’s Twitter are absent from the Mastodon event detection results.

\subsection{Coverage of selected keywords for tweet collection}

Due to the immense volume of daily tweets, collecting the entire tweet stream would result in excessive computational and storage costs. 
Following existing literature, we also use a set of keywords to filter out tweets potentially related to security.

As discussed in Section~\ref{sec:overall_workflow}, we obtained a keyword list from our collaborating Threat Intelligence (TI) company. 
A comparison of our keyword list with that used in previous work, W2E~\cite{shin2020cybersecurity}, revealed that 44.83\% of tweets collected using our keywords matched 83.1\% of the keywords from W2E. 
In other words, more than half of the tweets collected by our method could not be captured with the W2E keywords list. 
The missing W2E keywords in our list tend to be longer terms that are infrequently used on Twitter due to its post length limit, such as “distributed-denial-of-service” and “zero-day-exploit.”

Furthermore, we identified that essential keywords for certain categories, such as “phishing,” “cve,” and “patch,” were absent in the W2E keyword list. 
It indicates that our keyword list is more comprehensive, making our study offer a broader analysis of tweets compared to existing works.

\subsection{IDS/CTI engines}
Intrusion Detection Systems (IDS) and Cyber Threat Intelligence (CTI) engines, such as OpenCTI and Recorded Future, are established tools in the cybersecurity domain, each serving distinct roles. 
IDS systems are designed to monitor network traffic for suspicious activities and potential threats, while CTI engines aggregate and analyze threat data from various sources to provide actionable threat intelligence. 
\model{}, however, is designed not to replace these systems but to enhance the process of threat intelligence gathering.

\model{} can filter out tweets related to ongoing security events, focusing on pressing threat information. 
This refined information can then be utilized by existing enterprise IDS to extract Indicators of Compromise (IoCs), which are critical for detecting and preventing cybersecurity threats. 
Previous studies have demonstrated the effectiveness of capturing ongoing malware threats from IoC extracted from Twitter~\cite{twiti}. 
Furthermore, \model{} can be integrated into CTI engines to filter out irrelevant information, efficiently presenting TI-related tweets. 
Notably, some CTI companies have already incorporated social media into their TI feeds~\cite{recorded_TI_feed}

\subsection{STIX Integration}
Our method incorporates the identification of STIX objects during the tweet embedding process (Section~\ref{sec:emb_method}), which naturally sets the foundation for compatibility with existing STIX frameworks. 
The relationships between these objects are explicitly defined within the STIX 2.1 framework to form the STIX graph. 
By defining the relationships between these objects according to the STIX 2.1 framework, STIX graph~\cite{stix_relationsip} can be created. 
This graph facilitates the structured representation and analysis of threat information, enhancing the interoperability and utility of our system within the broader cybersecurity ecosystem~\cite{stix_usecase1, stix_usecase2}.

\subsection{Multi-lingual Support}
While our work focuses on tweets in English, our framework can be extended to other languages.
One straightforward approach is to translate tweets in other languages into English and then apply our proposed framework. 
This translation can be done effectively with the power of recent LLMs~\cite{llm_translation} 
However, this approach may introduce additional computational costs, particularly given the volume of tweets before categorization.

An alternative approach involves developing language-specific techniques for each stage as illustrated in Figure~\ref{fig:overview}. 
In the Tweet Category Tagging step, specific classifiers for each language need be developed. 
Thus, it is necessary to develop a dataset as in Section~\ref{sec:tweet_categorization}. 
Rather than building these models from scratch, we can leverage our existing dataset by translating the tweets into the target languages and then fine-tuning with additionally labeled tweets in those languages. 
This approach can significantly reduce the human effort required for dataset annotation in each language, which is a potential shortcoming of this approach.

For STIX object detection (NER) in the Tweet Embedding step, similar to the process for English, we can utilize the promptNER strategy by leveraging the multilingual capabilities of existing LLMs~\cite{bloom, gpt4_report}. Additionally, in the text feature engineering step, multilingual text embedding methods~\cite{multilingual_txtemb} can be employed. 
By incorporating these techniques, we can generate tweet embeddings for tweets in various languages. 
Consequently, the final step, Event Identification, can remain unchanged.

\subsection{Limitations and Future Work}
\label{sec:futurework}
\noindent \textbf{Security Events with Large Temporal Gaps:}
Our framework currently identifies each security event as a distinct occurrence if there is a significant temporal gap between related events. 
While this approach allows for a fine-grained analysis of events as they occur, in practice, linking related events over time can provide a more comprehensive understanding of threat evolution.

However, this limitation can be easily addressed through post-processing techniques that connect related events based on shared characteristics such as common CVE IDs or associated threat actors. 
As discussed in Section~\ref{sec:emb_method}-B, our method extracts STIX II objects from tweets, which can be utilized to identify and link these events. 
Security practitioners can enhance their analysis by associating events that share common threat actors, campaigns, or locations, thereby gaining a more integrated view of ongoing threats.

\noindent \textbf{Other important features}:
In this study, we focused on utilizing tweet content and temporal information to learn event attribute-centric embeddings.
However, Twitter offers several additional features that have the potential to enhance the performance of our model. 
These include hashtags, mentions, and retweets.
Hashtags are commonly used to highlight topics or important keywords in tweet content. 
While we removed hashtags in our study due to their overuse (especially by users aiming for broader exposure), it is worth noting that hashtags can provide valuable information for event detection if unrelated hashtags can be appropriately filtered. 
Mentions are used to include users related to the tweet content or to respond to specific users. 
Incorporating mentions can offer different aspects of tweet relationships, and thus can provide valuable contextual information.
Retweets refer to the action of sharing someone else's tweet with your own followers and can be useful for measuring a tweet's influence and popularity. 
By assigning more weight or importance to tweets with a higher number of retweets, we can enhance the effectiveness of our model in capturing influential tweets.
As part of our future work, we intend to incorporate these additional features into our model.

\subsection{Ethical Consideration}
\noindent \textbf{Use of Twitter API}.
This study utilizes the Twitter Enterprise API in full compliance with Twitter's privacy policies. 
To ensure the confidentiality of user data, we only collect and utilize tweets for research purposes. 
In line with Twitter's guidelines~\cite{twitter_privacy_policy}, which allows the sharing of Tweet IDs for research validation, we provide our dataset comprising solely of Tweet IDs. 

\noindent \textbf{Sensitive information masking}.
Recognizing the potential presence of private data in tweets, such as personal contact details or IP addresses, we have implemented robust data masking protocols. 
Prior to analysis, sensitive information within the tweets is masked to prevent our Pretrained Language Model (PLM) based Tweet Category Tagger from processing or learning from such data. 
This step is crucial in maintaining ethical standards and protecting individual privacy in our research methodology.

%% file: sections/7_relatedwork.tex
\section{Related Work}
\noindent\textbf{Twitter analysis for security}.
Twitter has been extensively analyzed for various security purposes. 
Shin et al.~\cite{twiti} focus on malware Indicators of Compromise (IOCs) on Twitter.
Through the collection and analysis of IOCs, they demonstrate Twitter's capability to capture ongoing malware threats. 
Sabottke et al.~\cite{sabottke2015vulnerability} conducted an in-depth investigation on vulnerability-related information on Twitter and showed that Twitter is able to identify more real-world vulnerabilities than public sources. 
Roy et al. analyze Twitter users' reports of phishing attacks~\cite{roy2021phishing} and malicious URLs~\cite{roy2021urls} on Twitter.
There are also some work that investigate fake accounts~\cite{mazza2022ready}, malware propagation~\cite{mal2013prop, mal2017prop}, and malware discovery on Twitter~\cite{mal2017discovery}. 
These studies utilize Twitter as a data source for identifying specific threats, while our studies focus on identifying any security-related events on Twitter. 
These security events encompass any occurrence or incident that can potentially impact the security of an
organization’s information technology (IT) systems, data, or overall cybersecurity posture.

\noindent\textbf{Security event detection}.
The domain of security event detection can be further categorized into two distinct categories. 
The first category is to detect malicious security events or activities through the analysis of system logs, syscall traces, and similar data sources.
Provenance graphs, which are constructed from auditing logs, are widely used for detecting malicious behaviors~\cite{hassan2019nodoze, hassan2020omegalog, han2020unicorn, ujcich2021causal}. 
Additionally, there are also some prior works dedicated to the automated correlation of these security events~\cite{van2022deepcase} and the prediction of future security events~\cite{shen2018tiresias, naseri2022cerberus}.

In another category, researchers focus on detecting emerging security events from social media, such as new data breaches, attacks, etc. 
Previous work employed machine learning techniques to classify tweets into predefined categories~\cite{ritter2015weakly, cydec, TwitterThreats}. 
However, these methods primarily categorize tweets and do not identify specific event instances.
Subsequently, Shin et al.~\cite{shin2020cybersecurity} and Le Sceller et al.~\cite{le2017sonar} employed keyword filtering and clustering techniques on tweet embeddings to identify specific instances of security events.
Nevertheless, these approaches relied on keywords, limiting their ability to capture contextual details. Furthermore, simply applying general text embedding methods to tweets for clustering may lead to less effective event identification (low event detection coverage) compared to dedicated embedding methods (shown in Section~\ref{sec:exp_emebdding}).

%% file: sections/8_conclusion.tex
\section{Conclusion}
The ever-evolving cybersecurity threat landscape requires security practitioners to stay current with the latest trends.
Although Twitter is acknowledged as a valuable and timely resource for information, conventional techniques for automatic security event detection on this platform have proven inadequate.
To fill this gap, our study introduces a security event attribution-centric tweet embedding method, which outperforms previous text and graph-based methods in effectiveness.
With this method, we have developed a security event detection framework, referred to as \model{}. 
This framework significantly improves performance, with the capability to identify twice as many security events as the baseline models.

%% file: sections/appendix.tex
\appendix

\subsection{Detailed Tweet Category Annotation Guideline}
In Section~\ref{sec:tweet_categorization}, we proposed a new multi-label tweet categorization dataset. 
The detailed definitions of each category, along with example tweets provided to the annotators. 
Additionally, following the consensus discussions, we established further criteria for tweet annotation. 
All of these details can be found in the released artifact.

\subsection{Tweet Category Tagging Performance}
\label{sec:multilable_performance}
\input{tables/classification_result} 
\input{tables/multi_label_result}

\noindent\textbf{Experimental Setup:}
To measure the performance, we partitioned data with an iterative stratification method from the scikit-multilearn\footnote{\url{http://scikit.ml}} package.
This approach ensures a balanced distribution of labels across the training, validation, and test subsets. 
For this experiment, 60\% of the data was allocated for training the Tweet Classifier, and the remaining portion was evenly divided into the validation (20\%) and test sets (20\%).
For base PLM, we tested with different variations of encoder language models, such as BERT~\cite{bert}, RoBERTa~\cite{roberta}, BERTweet~\cite{bertweet}, and SecureBERT~\cite{securebert}.

\noindent\textbf{Evaluation Metrics:}
Precision, recall, and F1-score are used for evaluating its classification performance.
Besides, the multi-label performance of our classifier is evaluated with five established metrics used for multi-label classification: Hamming Loss, Jaccard index, Subset accuracy, Micro-F1, and Macro-F1~\cite{zhang2013review, zhang2021multi, salawu2021large}.

\noindent \textbf{Results \& Discussion}: Table~\ref{tab:classification_result} presents the classification performance of each model on individual categories. 
The security-domain-specific model, SecureBERT, outperformed other models across multiple categories, possibly due to the security domain expertise learned by the model.
Additionally, classification performance on the ``DoS/DDoS'' and ``Vulnerability'' categories are comparatively better than other categories. 
This can be attributed to the presence of distinct keywords commonly associated with these categories. 
For instance, DDoS-related tweets are often marked by keywords like ``DDoS'' or ``take down'', making it easier to correctly classify such tweets.
On the other hand, the performance in the ``Data Privacy'' category is comparatively lower.
This is because the ``Data Privacy'' category encompasses a broader range of content, including discussions on data protection techniques, news of data breaches, or vulnerabilities that can lead to information disclosure.

The multi-label classification performance can be seen in Table~\ref{tab:multilabel_result}.
BERTweet outperforms other PLMs in these multi-label evaluation metrics, possibly due to its familiarity with tweet domain text.
Based on the overall observation, it is evident that domain-adapted Pretrained Language Models (PLMs), specifically SecureBERT and BERTweet, generally outperform the general-purpose PLMs. 
Considering our inputs are security tweets, SecureBERT and BERTweet are better at capturing and understanding the linguistic features and patterns specific to the text domain.
Given these two evaluation results, we select BERTweet in our proposed framework. 

\subsection{Security Named Entity Recognition (NER) Evaluation}
\label{sec:ner_eval}
In this work, we utilize prompt-based Named Entity Recognition (NER) 
approach to identify important security entities within tweets. 
These extracted entities are then used to construct a tweet relation graph. 
Although various NER models are available~\cite{spacy_ner, stanford_ner}, their application is restricted to predefined entity types.
Adapting these to security entity extraction requires the creation of new datasets and fine-tuning.  
Therefore, we use the prompt-based method~\cite{ashok2023promptner} to minimize human effort and computational resources. 

To validate the effectiveness of the prompt-based NER method in the security context, we perform an experiment on a cybersecurity NER benchmark dataset, CyNER~\cite{cyner}. 
We use OpenAI API with GPT-3.5-turbo, inputting prompts that include definitions of target entities and tweets. 
We focus on entities that overlap with the STIX objects we used for constructing a tweet relation graph.
The prompts are also designed to elicit structured responses, making it easy to extract entities and their corresponding entity type.

\input{tables/ner_result}

Table~\ref{tab:ner_result} presents the result in comparison to trained encoder-based NER models.
While our prompt-based approach adopted in our methodology does not surpass all other baseline methods, it outperforms RoBERTa models. 
Further analysis on false cases in prompt-based results reveals a notably low precision in the ``vulnerability'' entity type and low recall in the ``Operating system'' entity type.
This issue stems from the absence of publicly available detailed guidelines, leading to incorrect identification or omission of certain ambiguous entities. 
For example, the terms `malicious redirect' and ``vulnerabilities in BLU'' are classified as vulnerabilities by the prompt-based model, contrary to the original dataset's categorization.  
Also, certain entities like ``fake Netflix app'' and ``Google Play'' were not identified as `Operating system' by the prompt-based approach, although they are categorized as such in the original dataset. 
This discrepancy contributed to a lower recall in this category. Additionally, entities such as `Cenotix Python Keylogger' and `credential stealer' were not classified as `Malware' in the original dataset, which we suspect may be due to annotator mistakes.
Nevertheless, our prompt-based method remains effective for our purposes, given the consistent definitions we employ for extracting security objects from tweets.

%% file: tables/classification_result.tex
\begin{table*}[ht!]
\footnotesize
\ra{1.1}
\caption{
Tweet classification results. Only F1 score is reported in this table. 
}
\label{tab:classification_result}
\begin{adjustbox}{width=\textwidth, center}
\begin{tabular}{ccccccccc}
\toprule
& Non-Security & Uninformative & Data Privacy & Fraud/Phishing & Ransomware/Malware & DoS/DDoS & Vulnerability & Avg. F1 \\
\midrule
BERT       & 0.8978 & 0.7036 & 0.6940 & \textbf{0.7812} & \textbf{0.8796} & \textbf{0.9516} & 0.9159 & 0.8320 \\
RoBERTa    & 0.8892 & \underline{0.7125} & 0.7016 & \underline{0.7659} & 0.8577 & 0.9450 & 0.9179 & 0.8271 \\
BERTweet   & \textbf{0.9025} & 0.7108 & \underline{0.7202} & 0.7632 & 0.8614 & 0.9459 & \underline{0.9211} & \underline{0.8322} \\
SecureBERT & \underline{0.9021} & \textbf{0.7209} & \textbf{0.7278} & 0.7339 & \underline{0.8691} & \underline{0.9461} & \textbf{0.9334} & \textbf{0.8333}
\\
\bottomrule
\end{tabular}
\end{adjustbox}
\end{table*}

%% file: tables/multi_label_result.tex
\begin{table*}[ht]
% \footnotesize
\ra{1.1}
\caption{Multi-label classification results. $\uparrow$: higher the better; $\downarrow$: lower the better}
\label{tab:multilabel_result}
\begin{adjustbox}{width=.75\textwidth, center}
    \begin{tabular}{lccccc}
    \toprule
    & \makecell{Hamming loss ($\downarrow$)}
    & \makecell{Jaccard index ($\uparrow$)}
    & \makecell{Subset accuracy ($\uparrow$)}
    & Micro-F1 ($\uparrow$) 
    & Macro-F1 ($\uparrow$)\\
    \midrule
    BERT
    &  0.0489 & 0.8367 & 0.7547 & 0.8468 & \underline{0.8417} \\
    RoBERTa    
    & 0.0500 & 0.8319 & 0.7359 & 0.8448 & 0.8325 \\
    BERTweet
    & \textbf{0.0425} & \textbf{0.8565} & \textbf{0.7975} & \textbf{0.8633} & \textbf{0.8467} \\
    SecureBERT        
    & \underline{0.0447} & \underline{0.8494} & \underline{0.7783} & \underline{0.8584} & 0.8313 \\
    \bottomrule
    \end{tabular}
\end{adjustbox}
\vspace{-5pt}
\end{table*}

%% file: tables/ner_result.tex
\begin{table}[t!]
\caption{Security NER Performance}
\label{tab:ner_result}
\ra{1.1}
\begin{adjustbox}{width=.8\columnwidth, center}
\begin{tabular}{lccc}
\toprule
& \textbf{Precision} & \textbf{Recall} & \textbf{F1 Score} \\
\midrule
BERT-base-uncased & 69.67 & 69.88 & 69.77 \\
BERT-large-uncased & 72.69 & 73.45 & 73.07 \\
RoBERTa-base & 37.22 & 42.50 & 39.69 \\
RoBERTa-large & 34.76 & 44.18 & 38.91 \\
\midrule
Prompt-based & 41.85 & 62.68 & 50.19  \\ 
\bottomrule
\end{tabular}
\end{adjustbox}
\vspace{-10pt}
\end{table}